\documentclass[]{fairmeta}

\usepackage[utf8]{inputenc}
\usepackage{microtype}

\usepackage{subcaption}
\usepackage{caption}
\usepackage{wrapfig}
\usepackage{lineno}
\usepackage{subfiles}
\usepackage{graphicx}
\usepackage{booktabs}
\usepackage{diagbox}
\usepackage{float}
\usepackage{textcomp}
\usepackage{enumitem}
\usepackage{colortbl}
\usepackage{todonotes}
\usepackage{multirow}
\usepackage{subcaption}
\usepackage{amsmath}
\usepackage{dsfont}
\usepackage{amssymb}
\usepackage{listings}

\definecolor{mygreen}{HTML}{00A64F}
\usepackage{array}
\usepackage{lipsum}
\usepackage{longtable} 
\usepackage{adjustbox}
\usepackage[most]{tcolorbox}  
\usepackage{minted}           
\usepackage{wrapfig}
\usepackage{makecell}
\usepackage{rotating}
\usepackage{url}
\usepackage{threeparttable}
\usepackage{tabularx}
\newtcblisting{promptbox}[1][]{
    breakable,
    enhanced,
    coltitle=white,
    listing only,
    boxed title style={sharp corners, frame hidden},
    fonttitle=\ttfamily\centering\bfseries,
    title={#1},
    listing options={
        language={},
        basicstyle=\small\ttfamily,
        breaklines=true,
        breakautoindent=false,
        breakindent=0pt,
        columns=fullflexible,
        showstringspaces=false,
    },
    boxsep=2pt,
    top=2pt,
    bottom=2pt,
}

\newcommand{\Dataset}{\texttt{FRESCO}}

\newcommand{\rot}[1]{\rotatebox{90}{#1}}

\usepackage{xfp}
\usepackage[table]{xcolor}
\usepackage{colortbl}

\definecolor{MaxPerfColor}{HTML}{2B8CBE}

\newcommand{\highlightcell}[3]{%
  \begingroup
    \edef\percent{%
      \ifdim#2pt=#3pt
        0%
      \else
        \fpeval{round(max(0, min(100, ((#1)-(#2))/((#3)-(#2))*100)),0)}%
      \fi
    }%
    \xdef\__tvwikicolor{MaxPerfColor!\fpeval{0.8*\percent}!white}%
  \endgroup
  \cellcolor{\__tvwikicolor}#1%
}

\usepackage{bbm}
\usepackage{fvextra}

\DefineVerbatimEnvironment{PromptBlock}{Verbatim}{
  fontsize=\small,
  frame=single,
  breaklines=true,
  breakanywhere=true,
  breakautoindent=false,
  breaksymbolleft=,
  tabsize=2
}
\usepackage{inconsolata}
\usepackage[ruled,vlined]{algorithm2e}
\usepackage{cleveref}
\crefname{algocf}{alg.}{algs.}
\Crefname{algocf}{Algorithm}{Algorithms}

\title{FRESCO: Benchmarking and Optimizing Re-rankers for Evolving Semantic Conflict in Retrieval-Augmented Generation}
\author[1,2,*]{Sohyun An}
\author[1]{Hayeon Lee}
\author[1]{Shuibenyang Yuan}
\author[1]{Chun-cheng Jason Chen}
\author[2]{Cho-Jui Hsieh}
\author[1]{Vijai Mohan}
\author[1]{Alexander Min}

\affiliation[1]{Meta Superintelligence Labs}
\affiliation[2]{UCLA}

\contribution[*]{Work done at Meta}
\contribution[\dagger]{Joint last author}

\abstract{
Retrieval-Augmented Generation (RAG) is a key approach to mitigating the temporal staleness of large language models (LLMs) by grounding responses in up-to-date evidence. Within the RAG pipeline, re-rankers play a pivotal role in selecting the most useful documents from retrieved candidates. However, existing benchmarks predominantly evaluate re-rankers in static settings and do not adequately assess performance under evolving information---a critical gap, as real-world systems often must choose among temporally different pieces of evidence.
To address this limitation, we introduce \textbf{\Dataset{}} (\textbf{F}actual \textbf{R}ecency and \textbf{E}volving \textbf{S}emantic \textbf{CO}nflict), a benchmark for evaluating re-rankers in temporally dynamic contexts. By pairing recency-seeking queries with historical Wikipedia revisions, \Dataset{} tests whether re-rankers can prioritize factually recent evidence while maintaining semantic relevance. Our evaluation reveals a consistent failure mode across existing re-rankers: a strong bias toward older, semantically rich documents, even when they are factually obsolete.
We further investigate an instruction optimization framework to mitigate this issue. By identifying Pareto-optimal instructions that balance \emph{Evolving} and \emph{Non-Evolving Knowledge} tasks, we obtain gains of up to 27\% on Evolving Knowledge tasks while maintaining competitive performance on Non-Evolving Knowledge tasks.
}

\date{\today}
\correspondence{Sohyun An at \email{sohyun0423@cs.ucla.edu}}

\metadata[Code]{\url{https://github.com/facebookresearch/fresco}}

\begin{document}

\maketitle

\section{Introduction}
Large Language Models (LLMs) have achieved strong performance across a wide range of NLP tasks \citep{brown2020language, wei2021finetuned, bommasani2022opportunitiesrisksfoundationmodels, chowdhery2023palm, zhao2025surveylargelanguagemodels}, yet their practical reliability is often constrained by \textit{temporal staleness}---the mismatch between static pretraining corpora and a rapidly changing world. Retrieval-Augmented Generation (RAG) has therefore become a standard remedy, grounding LLM outputs in external and up-to-date evidence \citep{lewis2020retrieval, karpukhin2020dense, gao2023retrieval, izacard2023atlas, asai2024self}. Within the RAG pipeline, re-rankers act as a key gatekeeper: given a pool of retrieved candidates, they must prioritize the most useful evidence so that downstream generation remains factually correct \citep{nogueira2019passage, mxbai, bge_embedding, jina-reranker-v3, RankGPT, Rankvicuna, Rankzephyr}.

Despite this central role, the prevailing evaluation paradigm for re-ranking largely assumes a non-evolving information environment. Widely used benchmarks \citep{msmarco, trecdl, nq, beir} typically operationalize relevance through semantic overlap or topical alignment, implicitly treating the target information as \emph{fixed over time}. This assumption is frequently violated in real-world RAG deployments: when facts change, retrieval systems can surface multiple snapshots of ostensibly relevant evidence, some outdated and others reflecting the current state. For example, a query asking for the \emph{latest stable version of vLLM} may retrieve two release notes or changelogs---one reporting version 0.5.0 from several weeks ago and another reporting version 0.6.0 from two days ago. Both documents can appear highly relevant under conventional semantic matching, yet only the latter is chronologically valid for answering the query. We refer to this phenomenon as \emph{Evolving Semantic Conflict}: the re-ranker must choose among multiple semantically relevant candidates whose content conflicts due to temporal evolution.

Handling temporal discrimination---incorporating factual recency in addition to semantic relevance---is a practical requirement for robust RAG, particularly when multiple candidates are similarly relevant to the query. Without this capability, a re-ranker may systematically promote semantically informative but obsolete documents, causing the generator to produce confident yet incorrect answers. However, current evaluation regimes provide limited visibility into this failure mode, leaving it unclear whether existing re-rankers can reliably resolve conflicts induced by evolving knowledge without sacrificing relevance.

To bridge this gap, we introduce \textbf{\Dataset{}} (\textbf{F}actual \textbf{R}ecency and \textbf{E}volving \textbf{S}emantic \textbf{CO}nflict), a benchmark for evaluating re-rankers in settings where \emph{semantic relevance is necessary but not sufficient}. \Dataset{} pairs recency-seeking queries with historical \emph{Wikipedia} revisions to construct candidate sets in which passages remain topically aligned yet differ in factual recency. This design isolates the key challenge faced by real-world RAG systems: selecting evidence that is both relevant \emph{and} chronologically valid when semantically similar candidates disagree. Our evaluation identifies a consistent failure mode across existing re-rankers: they exhibit a strong \emph{semantic bias}, frequently preferring older, contextually dense documents over newer ones even when the older evidence is factually obsolete.

We further study a Pareto-based instruction optimization framework for LLM-based re-rankers, which take instructions alongside the query and candidate documents. The framework explicitly captures the trade-off between \emph{Evolving Knowledge} tasks (where recency is essential) and \emph{Non-Evolving Knowledge} tasks (where semantic relevance is typically sufficient), yielding a spectrum of Pareto-optimal instructions that practitioners can select based on deployment needs. Empirically, a Pareto-optimal instruction that emphasizes Evolving Knowledge improves performance on such tasks by up to 27\% while maintaining competitive results on Non-Evolving Knowledge tasks.

In summary, our contributions are as follows:
\begin{itemize}
\item We formalize \emph{Evolving Semantic Conflict} and introduce \textbf{\Dataset{}}, a benchmark that systematically evaluates re-rankers under temporally evolving information.
\item We identify a consistent failure mode in existing re-rankers, showing a strong bias toward semantically rich but factually obsolete documents.
\item We investigate a Pareto-based instruction optimization framework for LLM-based re-rankers, enabling controllable trade-offs between \emph{Evolving Knowledge} and \emph{Non-Evolving Knowledge} tasks.
\end{itemize}

\section{Related Work}

\paragraph{Re-ranking and Its Evaluation.}
Two-stage retrieval pipelines---retrieve followed by re-rank---are a standard approach in information retrieval and are widely used to support downstream applications such as RAG \citep{gao2023retrieval}. Re-rankers have progressed from BERT-style cross-encoders \citep{nogueira2019passage} to more recent LLM-based re-rankers \citep{Rankvicuna, Qwen3}, with performance commonly reported on benchmarks such as MS MARCO \citep{msmarco}, TREC DL \citep{trecdl}, Natural Questions \citep{nq}, and BEIR \citep{beir}. However, these benchmarks largely encode a temporally static notion of relevance: once labels are created, they are treated as invariant. As a result, they provide limited insight into whether a re-ranker can select chronologically valid evidence when multiple semantically relevant candidates reflect different points in time.

\paragraph{QA Benchmarks under Evolving Knowledge.}
Evolving, time-sensitive knowledge has motivated benchmarks such as RealtimeQA \citep{realtimeqa} and FreshQA \citep{freshllms}. These resources are valuable for measuring end-to-end QA performance under recency demands, but they are not designed to isolate the re-ranking decision itself. In particular, they typically evaluate final answer accuracy and do not provide controlled candidate pools containing competing, semantically relevant evidence with explicit temporal conflicts. This makes it difficult to perform a granular analysis of a re-ranker's ability to resolve \emph{Evolving Semantic Conflict}. In contrast, \Dataset{} leverages Wikipedia revision histories to construct timestamp-conditioned candidate pools, enabling scalable benchmarking with fine-grained control over temporal conflicts while keeping semantic relevance comparable across candidates.

\begin{figure*}[t]
\centering
    \includegraphics[width=1.0\linewidth]{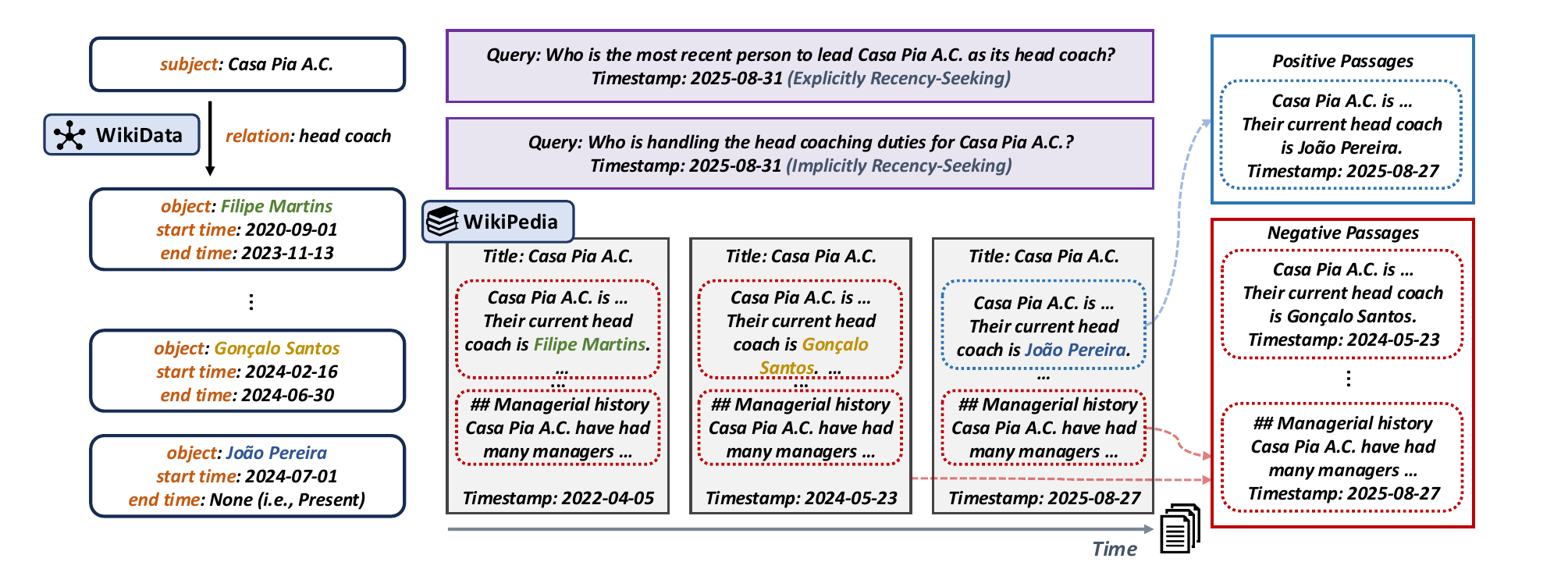}
    \vspace{-0.3in}
    \caption{
    \textbf{Illustration of the construction pipeline of \Dataset{}.} Time-annotated facts from Wikidata (left) are aligned with Wikipedia revisions (middle) to create positive and negative (outdated or irrelevant) passages (right) for explicit and implicit recency-seeking queries (top).
    }
    \label{fig:dataset}
\end{figure*}
\paragraph{Incorporating Temporal Signals in Ranking.}
A growing line of work incorporates temporal signals into ranking and retrieval by combining semantic relevance with recency-aware scoring, explicit time features, or modular symbolic components \citep{tempralm, freshllms, mrag}. While these approaches can improve performance on \emph{Evolving Knowledge} tasks, they often introduce an explicit recency preference that may be suboptimal when recency is irrelevant or misleading (\textit{e.g.}, \emph{Non-Evolving Knowledge} tasks). Our Pareto-based instruction optimization framework complements this direction by identifying a spectrum of Pareto-optimal instructions, enabling practitioners to choose operating points that balance Evolving and Non-Evolving Knowledge performance according to application needs.

\section{\Dataset{} Benchmark}
\label{sec:dataset_construction}

\Dataset{} is designed to evaluate re-rankers in controlled settings where semantic relevance alone is insufficient and temporal validity becomes necessary for selecting correct evidence. Each instance pairs a recency-seeking query with a candidate pool of passages that vary in semantic relevance and factual recency.
To enable scalable and reproducible construction, we leverage two complementary resources. \emph{Wikidata} provides structured facts as tuples with explicit temporal qualifiers (\textit{e.g.}, start and end dates), capturing how facts change over time. \emph{Wikipedia} revision histories then provide textual evidence corresponding to those facts at different points in time. By aligning these sources, we automatically generate benchmark instances without manual annotation. \Cref{fig:dataset} provides an overview of how \Dataset{} was constructed.

\subsection{Problem Formulation}
We represent a query as $q=(x_q, \tau_q)$, where $x_q$ is the query text and $\tau_q$ is the timestamp at which the query is posed. Each candidate passage is $c=(x_c, \tau_c)$, where $x_c$ is the passage text and $\tau_c$ is the document timestamp. To make temporal information available to the re-ranker, we append the timestamp to the textual input (\textit{e.g.}, \texttt{\{x\}\textbackslash nTimestamp: \{$\tau$\}}), where $\tau$ is formatted as \texttt{YYYY-MM-DDThh:mm:ssZ}. Given a candidate set $\mathcal{C}=\{c_1, c_2, \dots, c_n\}$, the goal is to rank highest the candidate that is (i) semantically relevant to $x_q$ and (ii) temporally valid at $\tau_q$.
Importantly, \Dataset{} includes candidates that are temporally valid but provide insufficient evidence to answer the query, as well as candidates that are semantically informative but factually outdated. Therefore, neither semantic relevance nor temporal validity alone is sufficient to solve the task.

\subsection{Construction Pipeline}
The construction proceeds in three phases: (1) fact identification and evidence alignment, (2) query generation and candidate pool construction, and (3) instance assembly via hard negative mining.

\paragraph{Phase 1: Fact Identification and Evidence Alignment.}
This phase identifies entities with verifiable temporal changes in Wikidata and aligns each fact with supporting evidence in Wikipedia. We first query Wikidata for relations with explicit temporal qualifiers (\textit{e.g.}, start time or end time). For each relation, we select entities with a sufficiently informative history, defined as having at least two distinct time-annotated facts between January 1, 2020, and a reference time $\tau_{\mathrm{ref}}$ (\textit{i.e.}, the query time assumed in our benchmark). Each fact is represented as a tuple $(s, r, o, t_s, t_e)$, where $s$ and $r$ denote the subject and relation, and $o$, $t_s$, and $t_e$ denote the object value and its validity period (start and end times). For a fixed subject--relation pair, the object value evolves over time; for example, for (Cristiano Ronaldo, member of sports team), $o$ changes from Juventus to Al Nassr with corresponding validity periods. This filtering ensures that selected entities exhibit meaningful temporal evolution rather than static attributes.
For each fact tuple, we consider its validity interval $[t_s, t_e)$ and retrieve Wikipedia page revisions for entity $s$ created during that interval. Because updates on Wikidata and Wikipedia are not always synchronized \citep{jang2022temporalwiki}, we apply a two-step verification procedure. First, we verify that the \emph{most recent} fact is supported by Wikipedia by parsing infoboxes from revisions in the corresponding validity interval and checking whether the target object value $o$ is explicitly stated; entities lacking such evidence are discarded. Second, we verify that \emph{at least one} earlier fact is also supported by Wikipedia by identifying an older revision whose infobox contains a different object value $o' \neq o$ corresponding to a previous state. This dual verification ensures that each retained entity is backed by textual evidence for both the latest fact and a prior conflicting fact, providing the necessary ingredients for \emph{Evolving Semantic Conflict}.

\paragraph{Phase 2: Query Generation and Candidate Pool Construction.}
After aligning verified temporal facts with Wikipedia revisions, we generate natural-language queries and construct corresponding candidate pools. For each subject ($s$)--relation ($r$) pair, we create a query targeting its most recent fact using manually crafted, relation-specific templates. To reflect real-world usage, we include both explicit and implicit recency-seeking forms. Explicit forms contain temporal keywords (\textit{e.g.}, ``Who is the \emph{most recent} person to lead \{$s$\} as its head coach?'') while implicit forms convey the same intent through tense (\textit{e.g.}, ``Who is handling the head coaching duties for \{$s$\}?''). The final query is obtained by filling \{$s$\} with the subject name (\textit{e.g.}, Tottenham Hotspur).
For each query, we construct a candidate pool of section-level passages extracted from the verified Wikipedia revisions. The positive passage is taken from the revision whose infobox matches the latest object value. We construct challenging negatives from two complementary sources. The first type consists of passages from the same (temporally valid) revision as the positive passage but drawn from sections that do not contain sufficient evidence to answer the query. The second type consists of outdated passages describing earlier facts, drawn from older revisions of the same entity and augmented with topically similar passages retrieved from a static Wikipedia snapshot (dated 2018-12-20) using a BGE retriever \citep{bge_embedding}. This design yields candidate pools containing passages that are semantically close to the query while introducing temporally conflicting evidence.

\paragraph{Phase 3: Instance Assembly via Hard Negative Mining.}
In the final phase, we assemble benchmark instances via hard negative mining. We consolidate all candidate negatives for a query into a single pool and score each passage using Qwen3-Embedding-0.6B \citep{Qwen3}. Each instance is represented as a triplet $(q, \mathcal{C}_q, \mathcal{R}_q)$, where $q$ is the query, $\mathcal{C}_q$ contains the positive passage and the top-$k$ highest-scoring negatives ($k=50$), and $\mathcal{R}_q$ provides binary labels (1 for the positive, 0 for negatives). This setup evaluates whether a re-ranker can prioritize evidence that is both relevant and temporally valid when faced with \emph{Evolving Semantic Conflict}. In total, \Dataset{} contains 3,658 queries per query type (7,316 queries overall), each paired with one positive passage and fifty negatives.

\begin{table*}[!t]
\centering
\vspace{-0.3in}
\caption{\textbf{\Dataset{} benchmark results for existing re-rankers.} Re-rankers are ordered by their release dates. Detailed model specifications are provided in \Cref{sec:appendix_model_details}. Cells are highlighted based on the numerical value normalized across each column, with darker blue indicating a higher performance.}
\label{tab:benchmark_full}
\resizebox{1.0\textwidth}{!}{%
\begin{tabular}{@{}ll|ccccccc|c@{}}
\toprule
&Model&MAP&MRR@5&MRR@10&nDCG@5&nDCG@10&Hit Rate@5&Hit Rate@10&Obsolete\%\\
\midrule
\multirow{19}{*}{\rot{\textit{Explicitly Recency-Seeking Queries}}}
&MonoT5
&\highlightcell{26.56}{19.01}{71.29}
&\highlightcell{21.72}{14.60}{69.87}
&\highlightcell{24.59}{17.06}{71.10}
&\highlightcell{26.28}{21.28}{75.69}
&\highlightcell{33.29}{27.79}{78.60}
&\highlightcell{41.03}{36.03}{103.93}
&\highlightcell{62.74}{57.06}{104.67}
&89.75
\\
&RankT5
&\highlightcell{38.85}{19.01}{71.29}
&\highlightcell{35.14}{14.60}{69.87}
&\highlightcell{37.37}{17.06}{71.10}
&\highlightcell{40.21}{21.28}{75.69}
&\highlightcell{45.60}{27.79}{78.60}
&\highlightcell{55.52}{36.03}{103.93}
&\highlightcell{72.17}{57.06}{104.67}
&90.24
\\
&UPR
&\highlightcell{35.69}{19.01}{71.29}
&\highlightcell{32.86}{14.60}{69.87}
&\highlightcell{34.34}{17.06}{71.10}
&\highlightcell{36.69}{21.28}{75.69}
&\highlightcell{40.47}{27.79}{78.60}
&\highlightcell{52.93}{36.03}{103.93}
&\highlightcell{64.00}{57.06}{104.67}
&86.18
\\
&RankGPT(gpt-3.5)
&\highlightcell{33.04}{19.01}{71.29}
&\highlightcell{30.98}{14.60}{69.87}
&\highlightcell{32.16}{17.06}{71.10}
&\highlightcell{41.16}{21.28}{75.69}
&\highlightcell{43.92}{27.79}{78.60}
&\highlightcell{71.16}{36.03}{103.93}
&\highlightcell{79.50}{57.06}{104.67}
&87.99
\\
&RankGPT(gpt-4o)
&\highlightcell{49.51}{19.01}{71.29}
&\highlightcell{49.38}{14.60}{69.87}
&\highlightcell{49.49}{17.06}{71.10}
&\highlightcell{62.30}{21.28}{75.69}
&\highlightcell{62.55}{27.79}{78.60}
&\highlightcell{98.93}{36.03}{103.93}
&\highlightcell{99.67}{57.06}{104.67}
&90.44
\\
&RankVicuna
&\highlightcell{38.14}{19.01}{71.29}
&\highlightcell{34.32}{14.60}{69.87}
&\highlightcell{36.48}{17.06}{71.10}
&\highlightcell{38.33}{21.28}{75.69}
&\highlightcell{43.62}{27.79}{78.60}
&\highlightcell{50.55}{36.03}{103.93}
&\highlightcell{67.09}{57.06}{104.67}
&89.29
\\
&RankZephyr
&\highlightcell{41.68}{19.01}{71.29}
&\highlightcell{38.22}{14.60}{69.87}
&\highlightcell{40.38}{17.06}{71.10}
&\highlightcell{43.01}{21.28}{75.69}
&\highlightcell{48.33}{27.79}{78.60}
&\highlightcell{57.57}{36.03}{103.93}
&\highlightcell{74.22}{57.06}{104.67}
&89.59
\\
&bce-reranker-base-v1
&\highlightcell{32.53}{19.01}{71.29}
&\highlightcell{28.04}{14.60}{69.87}
&\highlightcell{30.79}{17.06}{71.10}
&\highlightcell{32.92}{21.28}{75.69}
&\highlightcell{39.56}{27.79}{78.60}
&\highlightcell{47.81}{36.03}{103.93}
&\highlightcell{68.32}{57.06}{104.67}
&87.15
\\
&InRanker
&\highlightcell{53.18}{19.01}{71.29}
&\highlightcell{50.25}{14.60}{69.87}
&\highlightcell{52.04}{17.06}{71.10}
&\highlightcell{36.56}{21.28}{75.69}
&\highlightcell{43.02}{27.79}{78.60}
&\highlightcell{65.12}{36.03}{103.93}
&\highlightcell{78.49}{57.06}{104.67}
&91.20
\\
&mxbai-rerank-base-v1
&\highlightcell{27.22}{19.01}{71.29}
&\highlightcell{22.59}{14.60}{69.87}
&\highlightcell{25.17}{17.06}{71.10}
&\highlightcell{27.55}{21.28}{75.69}
&\highlightcell{33.80}{27.79}{78.60}
&\highlightcell{42.70}{36.03}{103.93}
&\highlightcell{62.06}{57.06}{104.67}
&89.72
\\
&Twolar
&\highlightcell{47.33}{19.01}{71.29}
&\highlightcell{44.17}{14.60}{69.87}
&\highlightcell{46.08}{17.06}{71.10}
&\highlightcell{48.63}{21.28}{75.69}
&\highlightcell{53.24}{27.79}{78.60}
&\highlightcell{62.11}{36.03}{103.93}
&\highlightcell{76.30}{57.06}{104.67}
&88.88
\\
&jina-reranker-v1-tiny-en
&\highlightcell{40.88}{19.01}{71.29}
&\highlightcell{37.45}{14.60}{69.87}
&\highlightcell{39.40}{17.06}{71.10}
&\highlightcell{41.96}{21.28}{75.69}
&\highlightcell{46.65}{27.79}{78.60}
&\highlightcell{55.63}{36.03}{103.93}
&\highlightcell{70.09}{57.06}{104.67}
&85.98
\\
&jina-reranker-v1-turbo-en
&\highlightcell{44.90}{19.01}{71.29}
&\highlightcell{41.85}{14.60}{69.87}
&\highlightcell{43.54}{17.06}{71.10}
&\highlightcell{46.23}{21.28}{75.69}
&\highlightcell{50.34}{27.79}{78.60}
&\highlightcell{59.46}{36.03}{103.93}
&\highlightcell{72.23}{57.06}{104.67}
&86.18
\\
&jina-reranker-v2
&\highlightcell{61.05}{19.01}{71.29}
&\highlightcell{59.17}{14.60}{69.87}
&\highlightcell{60.66}{17.06}{71.10}
&\highlightcell{65.20}{21.28}{75.69}
&\highlightcell{68.75}{27.79}{78.60}
&\highlightcell{83.38}{36.03}{103.93}
&\highlightcell{94.20}{57.06}{104.67}
&91.78
\\
&gte-multilingual-reranker-base
&\highlightcell{55.07}{19.01}{71.29}
&\highlightcell{52.68}{14.60}{69.87}
&\highlightcell{54.47}{17.06}{71.10}
&\highlightcell{58.77}{21.28}{75.69}
&\highlightcell{63.03}{27.79}{78.60}
&\highlightcell{77.17}{36.03}{103.93}
&\highlightcell{90.19}{57.06}{104.67}
&84.50
\\
&LdIR-Qwen2-reranker-1.5B
&\highlightcell{42.11}{19.01}{71.29}
&\highlightcell{38.65}{14.60}{69.87}
&\highlightcell{41.26}{17.06}{71.10}
&\highlightcell{45.80}{21.28}{75.69}
&\highlightcell{52.02}{27.79}{78.60}
&\highlightcell{67.50}{36.03}{103.93}
&\highlightcell{86.50}{57.06}{104.67}
&93.99
\\
&IncontextReranker
&\highlightcell{24.01}{19.01}{71.29}
&\highlightcell{19.60}{14.60}{69.87}
&\highlightcell{22.06}{17.06}{71.10}
&\highlightcell{26.86}{21.28}{75.69}
&\highlightcell{32.79}{27.79}{78.60}
&\highlightcell{48.66}{36.03}{103.93}
&\highlightcell{66.92}{57.06}{104.67}
&90.92
\\
&Qwen3-Reranker-0.6B
&\highlightcell{57.47}{19.01}{71.29}
&\highlightcell{55.28}{14.60}{69.87}
&\highlightcell{57.01}{17.06}{71.10}
&\highlightcell{60.80}{21.28}{75.69}
&\highlightcell{64.95}{27.79}{78.60}
&\highlightcell{80.48}{36.03}{103.93}
&\highlightcell{93.00}{57.06}{104.67}
&95.13
\\
&Qwen3-Reranker-8B
&\highlightcell{66.29}{19.01}{71.29}
&\highlightcell{64.87}{14.60}{69.87}
&\highlightcell{66.10}{17.06}{71.10}
&\highlightcell{70.69}{21.28}{75.69}
&\highlightcell{73.60}{27.79}{78.60}
&\highlightcell{88.35}{36.03}{103.93}
&\highlightcell{97.10}{57.06}{104.67}
&95.73
\\
\midrule
\multirow{19}{*}{\rot{\textit{Implicitly Recency-Seeking Queries}}}
&MonoT5
&\highlightcell{28.16}{19.29}{68.21}
&\highlightcell{23.68}{15.00}{66.40}
&\highlightcell{26.23}{17.38}{67.86}
&\highlightcell{28.36}{22.62}{72.20}
&\highlightcell{34.59}{28.36}{75.54}
&\highlightcell{43.28}{38.28}{102.10}
&\highlightcell{62.52}{57.52}{104.23}
&91.00
\\
&RankT5
&\highlightcell{37.24}{19.29}{68.21}
&\highlightcell{33.54}{15.00}{66.40}
&\highlightcell{35.68}{17.38}{67.86}
&\highlightcell{38.57}{22.62}{72.20}
&\highlightcell{43.78}{28.36}{75.54}
&\highlightcell{53.83}{38.28}{102.10}
&\highlightcell{69.98}{57.52}{104.23}
&90.11
\\
&UPR
&\highlightcell{39.44}{19.29}{68.21}
&\highlightcell{36.76}{15.00}{66.40}
&\highlightcell{38.16}{17.38}{67.86}
&\highlightcell{40.09}{22.62}{72.20}
&\highlightcell{43.62}{28.36}{75.54}
&\highlightcell{54.81}{38.28}{102.10}
&\highlightcell{65.31}{57.52}{104.23}
&87.08
\\
&RankGPT(gpt-3.5)
&\highlightcell{28.23}{19.29}{68.21}
&\highlightcell{25.66}{15.00}{66.40}
&\highlightcell{27.08}{17.38}{67.86}
&\highlightcell{34.81}{22.62}{72.20}
&\highlightcell{38.17}{28.36}{75.54}
&\highlightcell{62.08}{38.28}{102.10}
&\highlightcell{72.28}{57.52}{104.23}
&90.56
\\
&RankGPT(gpt-4o)
&\highlightcell{48.15}{19.29}{68.21}
&\highlightcell{47.79}{15.00}{66.40}
&\highlightcell{48.11}{17.38}{67.86}
&\highlightcell{60.60}{22.62}{72.20}
&\highlightcell{61.32}{28.36}{75.54}
&\highlightcell{97.10}{38.28}{102.10}
&\highlightcell{99.23}{57.52}{104.23}
&91.26
\\
&RankVicuna
&\highlightcell{38.85}{19.29}{68.21}
&\highlightcell{35.26}{15.00}{66.40}
&\highlightcell{37.18}{17.38}{67.86}
&\highlightcell{39.60}{22.62}{72.20}
&\highlightcell{44.36}{28.36}{75.54}
&\highlightcell{52.76}{38.28}{102.10}
&\highlightcell{67.74}{57.52}{104.23}
&89.95
\\
&RankZephyr
&\highlightcell{40.65}{19.29}{68.21}
&\highlightcell{37.08}{15.00}{66.40}
&\highlightcell{39.35}{17.38}{67.86}
&\highlightcell{42.11}{22.62}{72.20}
&\highlightcell{47.70}{28.36}{75.54}
&\highlightcell{57.41}{38.28}{102.10}
&\highlightcell{74.90}{57.52}{104.23}
&90.73
\\
&bce-reranker-base-v1
&\highlightcell{34.54}{19.29}{68.21}
&\highlightcell{30.52}{15.00}{66.40}
&\highlightcell{32.91}{17.38}{67.86}
&\highlightcell{35.06}{22.62}{72.20}
&\highlightcell{40.88}{28.36}{75.54}
&\highlightcell{48.93}{38.28}{102.10}
&\highlightcell{66.98}{57.52}{104.23}
&87.96
\\
&InRanker
&\highlightcell{56.82}{19.29}{68.21}
&\highlightcell{54.00}{15.00}{66.40}
&\highlightcell{55.67}{17.38}{67.86}
&\highlightcell{36.63}{22.62}{72.20}
&\highlightcell{43.00}{28.36}{75.54}
&\highlightcell{65.61}{38.28}{102.10}
&\highlightcell{78.08}{57.52}{104.23}
&90.21
\\
&mxbai-rerank-base-v1
&\highlightcell{33.18}{19.29}{68.21}
&\highlightcell{29.02}{15.00}{66.40}
&\highlightcell{31.47}{17.38}{67.86}
&\highlightcell{34.33}{22.62}{72.20}
&\highlightcell{40.31}{28.36}{75.54}
&\highlightcell{50.46}{38.28}{102.10}
&\highlightcell{69.05}{57.52}{104.23}
&90.60
\\
&Twolar
&\highlightcell{49.09}{19.29}{68.21}
&\highlightcell{46.11}{15.00}{66.40}
&\highlightcell{47.91}{17.38}{67.86}
&\highlightcell{50.81}{22.62}{72.20}
&\highlightcell{55.19}{28.36}{75.54}
&\highlightcell{65.04}{38.28}{102.10}
&\highlightcell{78.62}{57.52}{104.23}
&89.76
\\
&jina-reranker-v1-tiny-en
&\highlightcell{45.09}{19.29}{68.21}
&\highlightcell{41.94}{15.00}{66.40}
&\highlightcell{43.67}{17.38}{67.86}
&\highlightcell{45.75}{22.62}{72.20}
&\highlightcell{49.93}{28.36}{75.54}
&\highlightcell{57.24}{38.28}{102.10}
&\highlightcell{70.17}{57.52}{104.23}
&86.85
\\
&jina-reranker-v1-turbo-en
&\highlightcell{46.19}{19.29}{68.21}
&\highlightcell{43.22}{15.00}{66.40}
&\highlightcell{44.81}{17.38}{67.86}
&\highlightcell{47.14}{22.62}{72.20}
&\highlightcell{51.02}{28.36}{75.54}
&\highlightcell{58.94}{38.28}{102.10}
&\highlightcell{71.00}{57.52}{104.23}
&87.13
\\
&jina-reranker-v2
&\highlightcell{62.85}{19.29}{68.21}
&\highlightcell{61.21}{15.00}{66.40}
&\highlightcell{62.51}{17.38}{67.86}
&\highlightcell{67.20}{22.62}{72.20}
&\highlightcell{70.31}{28.36}{75.54}
&\highlightcell{85.27}{38.28}{102.10}
&\highlightcell{94.78}{57.52}{104.23}
&91.34
\\
&gte-multilingual-reranker-base
&\highlightcell{61.72}{19.29}{68.21}
&\highlightcell{59.88}{15.00}{66.40}
&\highlightcell{61.36}{17.38}{67.86}
&\highlightcell{65.73}{22.62}{72.20}
&\highlightcell{69.25}{28.36}{75.54}
&\highlightcell{83.30}{38.28}{102.10}
&\highlightcell{94.01}{57.52}{104.23}
&85.75
\\
&LdIR-Qwen2-reranker-1.5B
&\highlightcell{43.75}{19.29}{68.21}
&\highlightcell{40.56}{15.00}{66.40}
&\highlightcell{42.96}{17.38}{67.86}
&\highlightcell{47.76}{22.62}{72.20}
&\highlightcell{53.51}{28.36}{75.54}
&\highlightcell{69.63}{38.28}{102.10}
&\highlightcell{87.23}{57.52}{104.23}
&94.43
\\
&IncontextReranker
&\highlightcell{24.29}{19.29}{68.21}
&\highlightcell{20.00}{15.00}{66.40}
&\highlightcell{22.38}{17.38}{67.86}
&\highlightcell{27.62}{22.62}{72.20}
&\highlightcell{33.36}{28.36}{75.54}
&\highlightcell{50.66}{38.28}{102.10}
&\highlightcell{68.40}{57.52}{104.23}
&91.20
\\
&Qwen3-Reranker-0.6B
&\highlightcell{49.43}{19.29}{68.21}
&\highlightcell{46.66}{14.18}{66.40}
&\highlightcell{48.83}{17.38}{67.86}
&\highlightcell{52.09}{22.62}{72.20}
&\highlightcell{57.33}{28.36}{75.54}
&\highlightcell{75.07}{38.28}{102.10}
&\highlightcell{90.84}{57.52}{104.23}
&96.80
\\
&Qwen3-Reranker-8B
&\highlightcell{58.54}{19.29}{68.21}
&\highlightcell{56.67}{15.00}{66.40}
&\highlightcell{58.25}{17.38}{67.86}
&\highlightcell{63.52}{22.62}{72.20}
&\highlightcell{67.31}{28.36}{75.54}
&\highlightcell{84.31}{38.28}{102.10}
&\highlightcell{95.82}{57.52}{104.23}
&98.26
\\
\bottomrule
\end{tabular}%
}
\end{table*}

\subsection{Quality Validation}
To validate \Dataset{}, we conduct a human evaluation on 200 randomly sampled instances. Three trained annotators are shown four candidates: the pipeline-labeled positive passage and its three highest-scoring hard negatives retrieved by Qwen3-Embedding-0.6B. Annotators select the single passage that best supports the query at the given query timestamp, and may choose \texttt{None} if no passage provides sufficient evidence. We observe near-perfect inter-annotator agreement (Fleiss' $\kappa = 0.9689$), and the human majority vote matches our pipeline labels in 98.5\% of cases. These results indicate that \Dataset{} provides reliable supervision for evaluating re-rankers under temporally evolving information. Further details are provided in \Cref{sec:appendix_human_eval}.

\section{Benchmarking Re-rankers}

\paragraph{Setting.}
We evaluate 19 existing re-rankers on \Dataset{} to assess their performance under \emph{Evolving Semantic Conflict}. Re-rankers are ordered by release date in \Cref{tab:benchmark_full}. We report standard ranking metrics---MAP, MRR@$k$, nDCG@$k$, and Recall@$k$---for $k\in\{5,10\}$. We additionally report the \emph{Obsolete Ratio (\%)}: among the negative passages ranked above the positive, the proportion that are \emph{factually outdated} (\textit{i.e.}, describing an earlier state) rather than \emph{temporally valid but contextually insufficient}. A high Obsolete Ratio indicates that errors are dominated by selecting obsolete evidence over up-to-date evidence.

\paragraph{Analysis.}
As shown in \Cref{tab:benchmark_full}, \Dataset{} clearly differentiates existing re-rankers under \emph{Evolving Semantic Conflict}. Performance varies widely across models, with MAP ranging from 24.01 to 66.29 on explicitly recency-seeking queries and from 24.29 to 62.85 on implicitly recency-seeking queries. More recently released re-rankers generally perform better, although the trend is not strictly monotonic. Among the strongest models are RankGPT (gpt-4o), InRanker, jina-reranker-v2, gte-multilingual-reranker-base, and Qwen3-Reranker-8B, whereas earlier or smaller models such as MonoT5, RankT5, UPR, IncontextReranker, and mxbai-rerank-base-v1 tend to perform worse. Taken together, these results show that \Dataset{} is sufficiently discriminative to separate re-rankers with stronger temporal discrimination from those that rely more heavily on semantic similarity.

At the same time, the benchmark reveals a strikingly consistent failure mode across the model spectrum: the Obsolete Ratio remains high for nearly all re-rankers (84\%--98\%). This indicates that errors are rarely driven by topical irrelevance; instead, models often prefer semantically rich but outdated passages over temporally valid evidence. We attribute this pattern to a mismatch between the requirements of temporally dynamic RAG and prevailing training and evaluation paradigms, which mainly reward semantic overlap and relevance matching while providing limited supervision for temporal validity under evolving knowledge \citep{msmarco, trecdl, nq, beir}.

Qwen3-Reranker-8B, the strongest model on average, illustrates this tension especially clearly. Although it achieves the best overall effectiveness, it also exhibits one of the highest Obsolete Ratios in the benchmark, suggesting that its remaining errors stem less from semantic mismatch than from insufficient temporal discrimination among multiple semantically plausible candidates. We also find that this model is sensitive to how recency requirements are expressed: its MAP drops from 66.29 to 58.54 when the recency need is implicit rather than explicit, even though the underlying information need is unchanged. This sensitivity points to the potential of input optimization to better elicit temporal discrimination. In the next section, we therefore explore an input optimization framework that steers the re-ranker toward resolving temporal contradictions without degrading overall ranking quality.

\begin{table*}[t]
\centering
\small
\caption{\textbf{Performance comparison on Evolving Knowledge ($\mathcal{D}_{\text{EK}}$) and Non-Evolving Knowledge ($\mathcal{D}_{\text{NEK}}$) task.} Our instruction optimization method (Pareto Solution 1-4) discovers a Pareto front offering superior trade-offs compared to baselines. The trade-off plot is provided in \Cref{appendix:pareto_plot}.}
\small
\resizebox{1.0\textwidth}{!}{
\begin{tabular}{l|ccc|ccc}
\toprule
\multirow{2}{*}{\textbf{Method}} & \multicolumn{3}{c|}{$\mathcal{D}_{\text{EK}} \uparrow$} & \multicolumn{3}{c}{$\mathcal{D}_{\text{NEK}} \uparrow$} \\
\cmidrule(lr){2-4} \cmidrule(lr){5-7}
& MAP & MRR@10 & nDCG@10 & MAP & MRR@10 & nDCG@10 \\
\midrule
\textbf{Base Model (Qwen3-Reranker-8B)} & 62.41 & 62.17 & 70.45 & 60.93 & 77.73 & 79.68 \\
\midrule
\textbf{Temporal-Aware Models} & & & & & & \\
\hspace{1em} TempRALM \citep{tempralm} & 77.79 & 77.69 & 82.73 & 40.32 & 40.11 & 58.29 \\
\hspace{1em} FreshPrompt \citep{freshllms} & 57.64 & 57.68 & 66.26 & 61.36 & \textbf{79.46} & \textbf{80.80}  \\
\hspace{1em} MRAG \citep{mrag} & \underline{78.41} & \underline{77.88} & 80.85 & 59.18 & 76.65 & 78.63 \\
\midrule
\textbf{Fine Tuning} & & & & & & \\
\hspace{1em}Point-wise Finetuning & 76.31 & 76.18 & 81.52 & 61.54 & 78.28 & 80.03 \\
\hspace{1em}List-wise Finetuning & 67.91 & 67.73 & 74.95 & \underline{62.02} & 77.58 & 79.75 \\
\midrule
\textbf{Ours} & & & & & & \\
\hspace{1em}Pareto Solution 1 & \textbf{79.20} & \textbf{79.12} & \textbf{83.93} & 59.41 & 76.41 & 78.45 \\
\hspace{1em}Pareto Solution 2 & 77.67 & 77.58 & \underline{82.77} & 61.02 & 77.93 & 79.71 \\
\hspace{1em}Pareto Solution 3 & 72.51 & 72.38 & 78.68 & 61.51 & 78.20 & 79.97 \\
\hspace{1em}Pareto Solution 4 & 68.88 & 68.71 & 75.76 & \textbf{62.27} & \underline{79.00} & \underline{80.55} \\
\bottomrule
\end{tabular}
}
\label{tab:mitigation}
\end{table*}
\section{Pareto-Based Instruction Optimization for Re-rankers}
\label{sec:mitigation}
In this section, we first formalize the problem setting and then study input-level optimization for re-rankers. 
Because the instruction is the only input component fully under practitioner control, 
we propose an instruction optimization framework to steer re-ranker behavior.

\subsection{Problem Formulation}
An LLM-based re-ranker such as Qwen3-Reranker \citep{Qwen3} takes three inputs: an instruction prompt $p$, a query $q$, and a candidate passage $c \in \mathcal{C}_q$. Given model parameters $\theta$, the re-ranker $f_\theta$ produces a ranking permutation $\pi_{p,q} = f_{\theta}(p, q, \mathcal{C}_q)$.
We consider two task types: \emph{Evolving Knowledge} (EK) tasks, where temporal validity is critical, and \emph{Non-Evolving Knowledge} (NEK) tasks, where semantic relevance typically suffices. Let $\mathcal{D}_{\text{EK}}$ and $\mathcal{D}_{\text{NEK}}$ denote their respective distributions, where each data point $(q, \mathcal{C}_q, \mathcal{R}_q)$ consists of a query, candidate passages, and ground-truth labels.
An instruction optimized for $\mathcal{D}_{\text{EK}}$ may overemphasize temporal signals, potentially degrading performance on $\mathcal{D}_{\text{NEK}}$ where such signals are irrelevant. We therefore formulate instruction optimization as a bi-objective problem, seeking Pareto-optimal instructions that balance both task types.
For an instruction $p$ and a task distribution $\mathcal{D}$, we define the expected utility as:
\begin{equation}
\mathcal{J}_{\mathcal{D}}(p) =
\mathbb{E}_{(q, \mathcal{C}_q, \mathcal{R}_q) \sim \mathcal{D}}
\left[
\mathcal{U}\!\left(f_{\theta}(p, q, \mathcal{C}_q), \mathcal{R}_q\right)
\right],
\end{equation}
where $\mathcal{U}$ denotes a ranking quality metric such as MAP.
We seek a set of Pareto-optimal instructions $P^*$ that maximize the following bi-objective function:
\begin{equation}
\mathbf{F}(p) =
\left(
\mathcal{J}_{\mathcal{D}_{\text{EK}}}(p),
\mathcal{J}_{\mathcal{D}_{\text{NEK}}}(p)
\right).
\end{equation}
An instruction $p^* \in P^*$ is Pareto-optimal if there exists no other instruction that improves performance on one objective without degrading performance on the other.
The resulting Pareto front $P^*$ characterizes the trade-offs between EK and NEK tasks, enabling practitioners to select instructions that best match their application requirements.

\subsection{Optimization via Evolutionary Search}
Given that the search space $\mathcal{P}$ (\textit{i.e.}, the space of natural language instructions) is discrete and non-differentiable, we develop an evolutionary algorithm to navigate it. The algorithm iteratively refines a population of instructions using two text-based genetic operators.

\paragraph{Mutation.}
Inspired by \citet{APO}, our mutation operator refines an instruction by diagnosing ranking errors and correcting them. This process mimics gradient-based optimization but operates entirely in the textual domain.
Specifically, for a given instruction $p$ and query $q$, an error occurs when a negative passage $c'$ is ranked above the positive passage $c^*$. Let $\pi_{p,q}$ be the ranking permutation induced by $p$ for $(q,\mathcal{C}_q)$. The error set $\mathcal{E}(p)$ is defined as the set of tuples $(q, c^*, \mathcal{E}_q(p))$, where
\begin{equation}
\mathcal{E}_q(p) = \left\{ c' \in \mathcal{C}_q\setminus\{c^*\} \;:\; \pi_{p,q}(c') < \pi_{p,q}(c^*) \right\}.
\end{equation}
Given an instruction $p$, we derive textual gradients by analyzing its ranking errors on a sampled training batch.
Specifically, the gradient estimation operator $\mathcal{G}_{\text{estimate}}$ (\Cref{appendix:gradient-estimation}) takes as input the instruction $p$ and its associated error instances $\mathcal{E}(p)$, and produces a set of textual gradients $\mathbf{g}$, where each $g$ is a natural language critique describing a concrete failure pattern:
\begin{equation}
\mathbf{g} = \mathcal{G}_{\text{estimate}}(p, \mathcal{E}(p)).
\label{eq:estimate}
\end{equation}
Each textual gradient is then applied by the gradient application operator $\mathcal{G}_{\text{apply}}$ (\Cref{appendix:gradient-application}) to generate a mutated instruction that aims to correct the identified error:
\begin{equation}
p_{\text{mut}} = \mathcal{G}_{\text{apply}}(p, \mathcal{E}(p), g).
\label{eq:mutation}
\end{equation}
This process yields multiple candidate instructions per parent $p$, enabling local exploration of the instruction space guided by observed ranking failures.

\paragraph{Crossover.}
While mutation refines individual instructions based on their local errors, it does not leverage complementary strengths across different instructions within the population.
To address this, we introduce a crossover operator that combines instructions that excel on different objectives.
Given two parent instructions $p_A$ and $p_B$ sampled from the current population, we compare their performance on a sampled training batch across the two objectives (EK and NEK), and select pairs with complementary strengths (\textit{e.g.}, $p_A$ stronger on EK and $p_B$ stronger on NEK).
We then construct contrastive example sets by identifying instances on which the stronger instruction succeeds while the weaker instruction fails for each objective.
For example, $\mathcal{E}_{A \succ B}$ consists of EK instances where $p_A$ succeeds but $p_B$ fails, and $\mathcal{E}_{B \succ A}$ is defined analogously on NEK.
Conditioned on both parent instructions and their respective contrastive example sets, the crossover operator $\mathcal{X}$ (\Cref{appendix:crossover}) synthesizes a child instruction that integrates the objective-specific strengths of each parent:
\begin{equation}
p_{\text{cross}} = \mathcal{X}(p_A, p_B, \mathcal{E}_{A \succ B}, \mathcal{E}_{B \succ A}).
\label{eq:crossover}
\end{equation}

\paragraph{Overall Algorithm.}
Our instruction optimization algorithm proceeds iteratively, starting from an initial instruction population $P_0$.
At each evolutionary round $t$, the current population $P_t$ first undergoes an \emph{Expansion} phase, during which new candidate instructions are generated via mutation and crossover.
Specifically, mutation applies the composition $\mathcal{G}_{\text{apply}} \circ \mathcal{G}_{\text{estimate}}$ to refine individual instructions based on their training-batch errors, while crossover ($\mathcal{X}$) combines pairs of instructions with complementary strengths across objectives.
Together, these operators produce a set of newly generated instructions, denoted $P_{\text{new}}$.
In the subsequent \emph{Evaluation} phase, each instruction $p$ in the combined pool $P_t \cup P_{\text{new}}$ is evaluated on a held-out validation minibatch to estimate its objective vector $\hat{\mathbf{F}}(p)$.
The \emph{Selection} phase then constructs the next generation $P_{t+1}$ by identifying the empirical Pareto front of non-dominated instructions.
If the Pareto front exceeds the budgeted population size, it is pruned using a diversity-preserving criterion such as crowding distance.
After a fixed number of evolutionary rounds, the algorithm terminates and returns the final Pareto front, providing a set of optimized instructions that expose different trade-offs between EK and NEK tasks.
The complete procedure is summarized in \Cref{sec:appendix_algo}.

\subsection{Experiments}
\label{sec:mitigation-experiments}

\begin{wraptable}{r}{0.18\textwidth}
\vspace{-0.8em}
\centering
\small
\begin{tabular}{lcc}
\toprule
Inst. & $\tau_q$ & $\tau_d$ \\
\midrule
Base & 5.56 & 31.70 \\
Opt. & \textbf{5.83} & \textbf{33.40} \\
\bottomrule
\end{tabular}
\caption{\textbf{Temporal Attention Ratio for Base vs. Optimized Instructions.}}
\label{tab:attention_wrap}
\vspace{-0.8em}
\end{wraptable}

\paragraph{Setting.}
We instantiate the two task distributions as follows: $\mathcal{D}_{\text{EK}}$ corresponds to our proposed \Dataset{} benchmark, while $\mathcal{D}_{\text{NEK}}$ is derived from a reformulated version of the NQ \citep{nq} dataset in which each query and passage is assigned a randomly generated timestamp. We maintain a fixed population size of 4 throughout all experiments. Full implementation details are provided in \Cref{sec:appendix_implementation_details}.
We compare our method against three categories of baselines.
\begin{wrapfigure}{r}{0.28\columnwidth}
\vspace{-10pt}
\centering
\includegraphics[width=\linewidth]{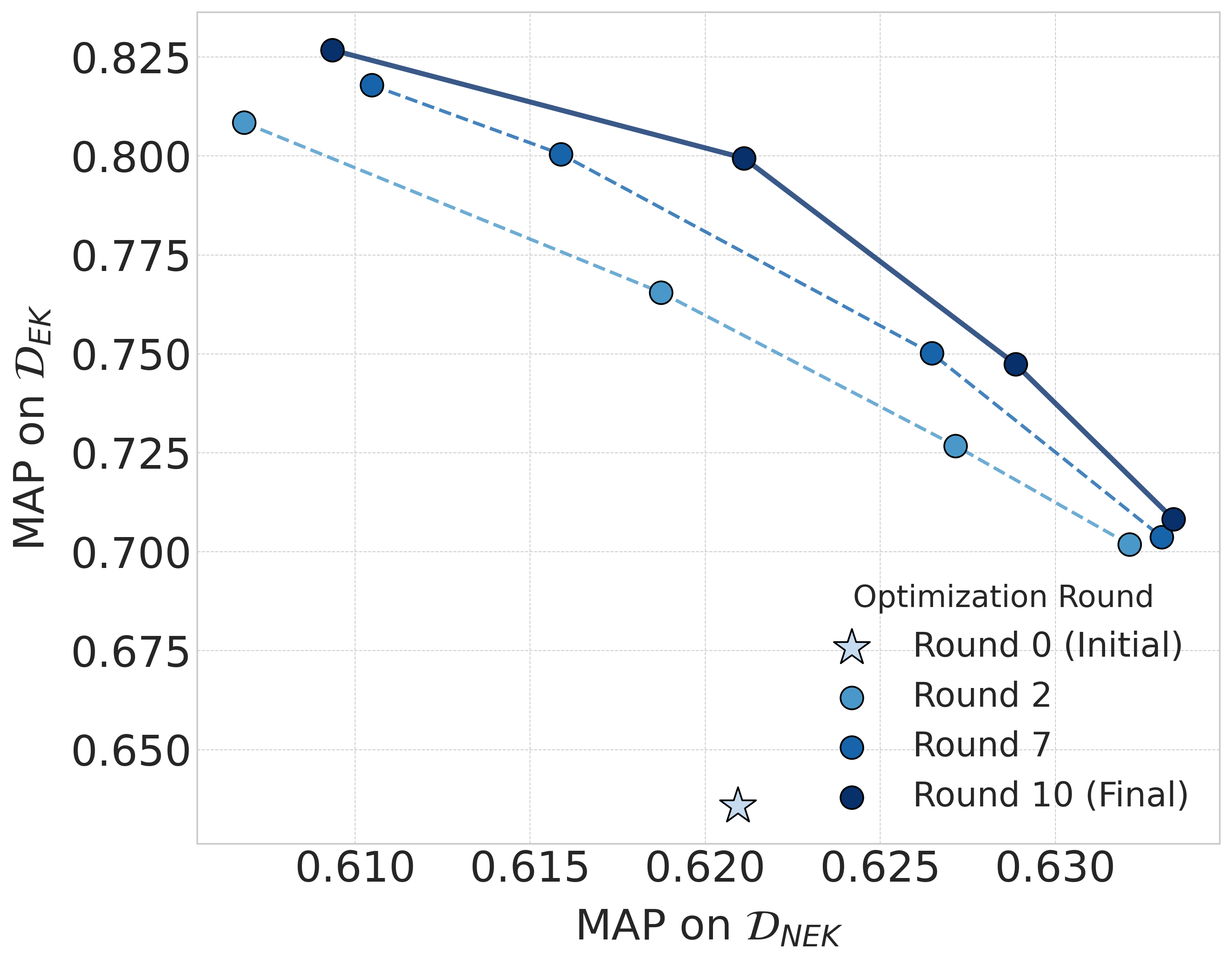}
\vspace{-20pt}
\caption{\textbf{Evolution of the Pareto front.} Starting from the initial instruction (star), optimization progressively improves the trade-off between $\mathcal{D}_\text{EK}$ and $\mathcal{D}_\text{NEK}$.}
\label{fig:pareto_evolution}
\end{wrapfigure}

\textbf{(i) Base Model.} We evaluate the off-the-shelf Qwen3-Reranker-8B \citep{Qwen3} using its default general-purpose instruction, representing standard re-ranking performance without temporal adaptation.
\textbf{(ii) Temporal-Aware Methods.} We include recent methods designed for handling temporal signals: \textbf{TempRALM} \citep{tempralm}, which augments semantic scores with temporal scores; \textbf{FreshPrompt} \citep{freshllms}, which leverages few-shot in-context learning for adaptation; and \textbf{MRAG} \citep{mrag}, which employs a hybrid semantic-temporal module-based approach.
\textbf{(iii) Fine-Tuning Baselines.} As supervised comparisons, we fine-tune the base model on a mixture of $\mathcal{D}_{\text{EK}}$ and $\mathcal{D}_{\text{NEK}}$ training data using standard point-wise and list-wise ranking losses.
Additional details are provided in \Cref{sec:appendix_baselines}.

\vspace{-10pt}
\paragraph{Results.}
As shown in \Cref{tab:mitigation}, our instruction optimization identifies a Pareto front that offers improved trade-offs over single-instruction baselines.
At one extreme, Pareto Solution 1 prioritizes $\mathcal{D}_{\text{EK}}$, achieving the highest MAP of 79.20 (a 27\% relative gain over the base model).
This outperforms strong baselines such as point-wise fine-tuning and MRAG on $\mathcal{D}_{\text{EK}}$, while avoiding the severe degradation on $\mathcal{D}_{\text{NEK}}$ observed with TempRALM.
At the other extreme, Pareto Solution 4 achieves the highest MAP on $\mathcal{D}_{\text{NEK}}$ (62.27) 
and remains competitive with the strongest baselines under MRR and nDCG.
Notably, it still achieves a MAP of 68.88 on $\mathcal{D}_{\text{EK}}$, improving over the base model despite prioritizing $\mathcal{D}_{\text{NEK}}$.
The intermediate solutions (Pareto Solutions 2 and 3) demonstrate a smooth and controllable trade-off between the two objectives.
Overall, our framework yields a set of specialized instructions that jointly dominate the trade-off frontier, enabling practitioners to select re-ranking behavior aligned with deployment priorities. We report results on additional re-rankers in \Cref{appendix:mitigation_rankllama}.

\begin{wraptable}{r}{0.52\columnwidth}
\centering
\small
\caption{\textbf{Ablation study.} Pareto Solutions 1 and 4 are the two extremes of the Pareto front, capturing opposite trade-offs between $\mathcal{D}_{\text{EK}}$ and $\mathcal{D}_{\text{NEK}}$. We report MAP on $\mathcal{D}_{\text{EK}}$ and $\mathcal{D}_{\text{NEK}}$.}
\begin{tabular}{l|c|c}
\toprule
Method & $\mathcal{D}_{\text{EK}}$ & $\mathcal{D}_{\text{NEK}}$ \\
\midrule
Pareto Solution 1 & 79.20 & 59.41 \\
Pareto Solution 4 & 68.88 & 62.27 \\
\midrule
Pareto Solution 1 (w/o Mut.) & 76.59 & 60.53 \\
Pareto Solution 4 (w/o Mut.) & 65.26 & 62.44 \\
\midrule
Pareto Solution 1 (w/o Cross.) & 77.09 & 59.43 \\
Pareto Solution 4 (w/o Cross.) & 62.60 & 62.23 \\
\bottomrule
\end{tabular}
\vspace{-10pt}
\label{tab:ablation}
\end{wraptable}

\paragraph{Analysis.}
To better understand the effects of our instruction optimization framework, we conduct additional analyses. An attention analysis, detailed in \Cref{appendix:attention} and summarized in \Cref{tab:attention_wrap}, shows that the instruction corresponding to the $\mathcal{D}_{\text{EK}}$-optimal point on the Pareto front (Pareto Solution 1) increases the re-ranker's attention to both query ($\tau_q$) and document timestamps ($\tau_d$), with a larger gain on $\tau_d$. This shift offers a plausible explanation for its improved temporal discrimination.
Moreover, we visualize the optimization dynamics in \Cref{fig:pareto_evolution}, which shows that the evolutionary process progressively expands the Pareto front across rounds, yielding instructions with improved trade-offs between $\mathcal{D}_{\text{EK}}$ and $\mathcal{D}_{\text{NEK}}$; additional details are provided in \Cref{appendix:opt_process}.

\paragraph{Ablation Study.}
We evaluate the impact of mutation and crossover by removing each operator while keeping the total instruction budget fixed.
Pareto Solutions 1 and 4 correspond to the $\mathcal{D}_{\text{EK}}$- and $\mathcal{D}_{\text{NEK}}$-optimal points on the Pareto front, respectively.
As shown in \Cref{tab:ablation}, removing mutation consistently reduces MAP on $\mathcal{D}_{\text{EK}}$ across solutions.
Removing crossover also degrades performance, with a larger drop in $\mathcal{D}_{\text{EK}}$ MAP for Pareto Solution 4, while MAP on $\mathcal{D}_{\text{NEK}}$ is comparatively less affected.
Overall, both operators contribute to the final Pareto front, and ablating either one yields inferior trade-offs between $\mathcal{D}_{\text{EK}}$ and $\mathcal{D}_{\text{NEK}}$.

\section{Conclusion}

We introduce \Dataset{} to evaluate re-rankers under \emph{Evolving Semantic Conflict}, where they must prioritize up-to-date evidence while preserving semantic relevance. Our analysis reveals a common failure mode: existing re-rankers often prefer semantically rich but obsolete evidence. To address this, we propose a bi-objective instruction optimization framework that yields a Pareto front over EK and NEK.



\section*{Limitations}
\label{sec:limitation}

This work focuses on a practically important aspect of re-ranking under temporally evolving information, and several considerations define the scope of the current study.
First, \Dataset{} is constructed using Wikipedia and Wikidata, which offer well-structured and verifiable records of factual evolution. This choice enables controlled and scalable benchmark construction, while other information sources, such as news or rapidly changing web content, may exhibit different temporal characteristics that are not explicitly captured here.
Second, the evaluation setup makes document timestamps explicitly available to the re-ranker. This design choice allows us to directly assess temporal discrimination in a controlled setting, but does not address scenarios where temporal signals are implicit, noisy, or unavailable and must instead be inferred from content.
Finally, instruction optimization operates solely at the interface level without modifying model parameters. This design enables lightweight applications even in black-box settings and complements model-level approaches that address temporal reasoning.
Overall, these considerations reflect the scope of the present study and highlight directions for extending temporally robust re-ranking evaluation and optimization in future work.

\section*{Ethical Considerations}
\label{sec:ethics}

This work includes a limited amount of human annotation conducted solely for evaluation purposes. The annotation task involves verifying the semantic relevance and temporal validity of publicly available textual passages and does not require subjective judgment or involve sensitive topics. The study was conducted in accordance with the authors’ institutional guidelines for research ethics and was determined to be exempt from formal ethics review.
The benchmark is constructed from content publicly available on Wikipedia and Wikidata. These resources are used in a manner consistent with their role as open encyclopedic knowledge bases. When using existing datasets and models, we adhere to their respective licenses and terms of use.
The artifact introduced in this work, \Dataset{}, is intended for academic research and benchmarking of re-rankers under temporally evolving information. Its intended use is compatible with the access conditions of the original data sources from which it is derived. The dataset is accompanied by usage guidelines that prohibit harmful or unethical applications, including deception, discrimination, or other harmful behaviors.
The dataset does not intentionally collect or include information about private individuals. The source materials consist of curated encyclopedic content describing public entities and factual information. During dataset construction, we checked the collected data and did not identify explicit offensive or harmful content. The dataset primarily covers English-language encyclopedic content and focuses on entities with temporally evolving factual attributes.


\bibliographystyle{assets/plainnat}
\bibliography{custom}

\clearpage
\newpage
\beginappendix

\appendix


\section{Details of Operators}

\subsection{Gradient Estimation Operator}
\label{appendix:gradient-estimation}
\begin{promptbox}[]
# Role and Goal

You are an expert Prompt Engineer specializing in optimizing inputs for ranking and information retrieval models. Your objective is to analyze the provided examples of failure from my current reranker model, diagnose the weaknesses in its current prompt, and craft superior, revised prompts that will improve its performance.

# Model Context

I am using a reranker model that takes three inputs: a **`Prompt`**, a **`Query`**, and a **`Document`**.

The model's function is to evaluate how relevant the `Document` is to the `Query` based on the guidance provided in the `Prompt`. It returns a relevance score, where a higher score indicates higher relevance.

# Current Prompt and Problem Statement

The current prompt I am using is:
`"{current_prompt}"`

For a given `Query`, the model often assigns a higher score to an irrelevant document (**Negative Document**) than to the ideal, relevant document (**Positive Document**). My goal is to fix this by improving the prompt.

# Error Examples for Analysis

Below are concrete examples where the current model fails. In each case, a `Negative Document` incorrectly receives a higher score than the `Positive Document`. Please analyze these patterns to identify the flaw in the current prompt.

{error_string}

# Core Task & Required Output

Give {num_gradients} reasons why the current prompt could have gotten these examples wrong. Wrap each reason with <START> and <END>
\end{promptbox}

\subsection{Gradient Application Operator}
\label{appendix:gradient-application}
\begin{promptbox}[]
# Role and Goal

You are an expert Prompt Engineer tasked with improving a reranker model's performance. The reranker model's job is to score how relevant a **`Document`** is to a **`Query`**, **based on the prompt I provide**. Your objective is to generate new, improved prompts for the reranker model based on a provided analysis of a failing prompt and its error examples.

# Background Information

You will be given three pieces of information: the current prompt that performed poorly, the specific examples it failed on, and an analysis of why it failed.

### 1. The Current Prompt

This is the prompt that needs improvement.
`"{current_prompt}"`

### 2. Error Examples

The current prompt failed on the following examples:
{error_str}

### 3. Analysis of the Problem

Based on the errors, the key weaknesses of the current prompt were identified as follows:
{gradient_str}

# Core Task & Required Output

Based on all the information provided above (the original prompt, its errors, and the analysis), please perform the following tasks:

1.  **Generate Prompts:** Write **{steps_per_gradient}** different and improved prompts that aim to overcome the identified weaknesses.
2.  **Encourage Diversity:** Each prompt should be distinct from the others.
3.  **Formatting:** Wrap each new prompt individually with `<START>` and `<END>`.
\end{promptbox}

\subsection{Crossover Operator}
\label{appendix:crossover}
\begin{promptbox}[]
# Role and Goal

You are an expert Prompt Engineer specializing in synergistic prompt design. Your objective is to analyze two distinct prompts, identify the core reasons for their unique successes, and then synthesize these insights into a superior, hybrid prompt that inherits the strengths of both.

# Contrastive Analysis of Two Prompts

We have analyzed two prompts, A and B, and have found specific examples where one succeeded while the other failed. This contrastive analysis reveals the unique strengths of each.

### 1. Prompt A's Strengths (Where A Succeeded and B Failed)

**Prompt A:** `"{prompt_a}"`

In the following examples, **Prompt A correctly identified the Positive Document, whereas Prompt B failed to do so.** These examples highlight the winning strategy of Prompt A.
{examples_a_wins}

### 2. Prompt B's Strengths (Where B Succeeded and A Failed)

**Prompt B:** `"{prompt_b}"`

In the following examples, **Prompt B correctly identified the Positive Document, whereas Prompt A failed to do so.** These examples highlight the winning strategy of Prompt B.
{examples_b_wins}

# Core Task & Required Output

Your task is to create a new, more powerful prompt by combining the winning strategies of both A and B.

1.  **Analyze Prompt A's Winning Strategy:** Based on the first set of examples, what specific phrasing, instruction, or principle in Prompt A allows it to succeed where B fails?
2.  **Analyze Prompt B's Winning Strategy:** Similarly, based on the second set of examples, what is the core strength of Prompt B that allows it to handle cases that A could not?
3.  **Generate Hybrid Prompts:** Synthesize these two winning strategies into **{num_crossovers}** distinct, new prompts. Each new prompt must be a cohesive instruction set that aims to solve all provided examples by intelligently combining the best of A and B.
4.  **Formatting:** Wrap each new prompt individually with `<START>` and `<END>`.
\end{promptbox}

\section{Additional Experiments}
\label{sec:appendix_additional_exp}

\subsection{How Instructions Steer Temporal Awareness of Re-rankers}
\label{appendix:attention}

Our primary results show that our instruction optimization method identifies a Pareto front capturing trade-offs between $\mathcal{D}_{\text{EK}}$ and $\mathcal{D}_{\text{NEK}}$. 
In particular, the $\mathcal{D}_{\text{EK}}$-optimal point (Pareto Solution 1) yields substantial performance gains on $\mathcal{D}_{\text{EK}}$. 
To better understand the mechanism underlying this improvement, we analyze how the optimized instruction alters model behavior. 
We hypothesize that it encourages the model to attend more strongly to explicit timestamp signals in both the query and the document.

\paragraph{Experimental Setup}
To test this hypothesis, we measure the \textbf{Temporal Attention Ratio}, defined as the proportion of attention that the re-ranker's final token directs towards the timestamp tokens within the input. The final token's attention distribution is a strong proxy for the model's decision process, as it directly precedes the final relevance judgment (\textit{e.g.}, \texttt{yes} or \texttt{no} tokens) \citep{Qwen3}.
Formally, let $\mathbf{A}^{(L)}$ be the attention matrix of the model's final layer $L$, and let $T_{\text{last}}$ be the index of the final token before producing the final relevance judgment. For a given passage, let $I_{\text{ts}}$ be the set of token indices corresponding to its timestamp (\textit{e.g.}, Timestamp: 2025-08-31T00:00:00Z), and let $I_{\text{context}}$ be the set of indices for all tokens from the start of the query to the end of the document. The Temporal Attention Ratio is calculated as:
\begin{equation}
    \text{Temporal Attention Ratio} = \frac{\sum_{i \in I_{\text{ts}}} \mathbf{A}^{(L)}_{T_{\text{last}}, i}}{\sum_{j \in I_{\text{context}}} \mathbf{A}^{(L)}_{T_{\text{last}}, j}}
\end{equation}
We compute this ratio for both the query's timestamp ($\tau_q$) and the document's timestamp ($\tau_d$). We compare two instructions: a generic \textbf{Base Instruction} (``Given a web search query, retrieve relevant passages that answer the query'') and our best-performing \textbf{Optimized Instruction} (Pareto Solution 1 from our Pareto front: ``Given a specific query, directly and accurately answer the question by retrieving the most recent, precise, and relevant document that provides the latest available data as of the query's timestamp. Ensure the document not only offers the most current information but also presents the answer in a clear and concise manner, reflecting the latest developments. Prioritize documents that are both reliable and up-to-date, filtering out outdated information to provide the most accurate response possible''). The analysis is performed on 1000 randomly sampled instances from our \Dataset{} test set using the Qwen3-Reranker-8B model.

\paragraph{Results and Analysis}
The results, summarized in \Cref{tab:attention_wrap}, reveal a clear shift in the re-ranker's attention patterns induced by our optimized instruction. We observe an increase in the attention paid to the timestamps in both the query and the document, with the most pronounced change occurring for the document's timestamp.

Specifically, with the Optimized Instruction, the average attention directed at the document's timestamp ($\tau_d$) increases from 31.70\% to 33.40\%. This notable shift suggests that the re-ranker, guided by the explicit instruction to prioritize recency, learns to more actively verify the temporal validity of the candidate passage when making its relevance judgment.
Furthermore, we observe a smaller but consistent increase in attention towards the query's timestamp ($\tau_q$), which rises from 5.56\% to 5.83\%. This suggests the optimized instruction enhances the re-ranker's overall sensitivity to the temporal context of the task. 

This analysis provides a mechanistic explanation for the performance gains reported in our main results (\Cref{tab:mitigation}). The instruction optimization is not a black box; it works by tangibly re-directing the re-ranker's internal attention mechanisms to focus on the crucial temporal cues required for the task. This confirms that our method effectively instills a temporal sensitivity into the re-ranker by reshaping its information processing strategy.

\subsection{Dynamics of Instruction Optimization Process}
\label{appendix:opt_process}

To gain insight into how our evolutionary search navigates the instruction space, we visualize the progression of the Pareto front over optimization rounds. \Cref{fig:pareto_evolution} shows the MAP of non-dominated instructions discovered over successive optimization rounds on the validation sets of $\mathcal{D}_{\text{EK}}$ and $\mathcal{D}_{\text{NEK}}$. Each point on a curve represents the averaged coordinates of a specific Pareto solution across 3 independent runs, and the lines connect these average points to illustrate the shape of the averaged front at that round.

The optimization begins at Round 0 with a single, general-purpose instruction (grey star), establishing the baseline performance. As the search progresses (represented by progressively darker blue points and dashed lines), we observe a clear upward and rightward shift of the Pareto front. This indicates that our algorithm, leveraging mutation via textual gradients and instruction crossover, successfully discovers new instructions that significantly improve upon the initial instruction, achieving better performance on one or both objectives simultaneously. 
By Round 10, the final averaged Pareto front (dark navy line and circles) represents the culmination of the optimization process. It clearly dominates the fronts from earlier rounds, signifying performance gains. The final front spans a considerable range, demonstrating that our method identifies multiple high-performing instructions: some specialized for $\mathcal{D}_{\text{EK}}$ (top-left region), others for $\mathcal{D}_{\text{NEK}}$ (bottom-right region), and several offering balanced performance in between.

These dynamics confirm the effectiveness of our evolutionary approach. The algorithm progressively explores the instruction landscape, escaping the limitations of the initial instruction and converging towards a set of diverse, Pareto-optimal instructions that effectively balance the competing demands of temporal awareness and general relevance.

\subsection{Case Study: Evolutionary Instruction Optimization Dynamics}
To provide a concrete illustration of our evolutionary instruction optimization framework, \Cref{tab:case-study} presents an example optimization trajectory. The table traces how a re-ranker instruction evolves across multiple rounds through mutation and crossover guided by feedback signals.

The initial instruction in Round~0 is adopted from the technical report of Qwen3-Reranker \citep{Qwen3} and represents a commonly used, high-level formulation for passage re-ranking. In Round~1, the gradient exposes limitations of this generic instruction, and mutation is applied to generate a modified variant. Since the initial population contains only a single instruction ($|P_0|=1$), crossover is not applicable at this stage.

As optimization proceeds, further mutation steps produce additional instruction variants. Once multiple candidates are available, crossover is applied to recombine elements from different parent instructions. As shown in Round~4, this process expands the set of candidate instructions explored during optimization.

Overall, this case study illustrates how mutation and crossover are used to explore a Pareto set of instructions across both Evolving Knowledge (EK) and Non-Evolving Knowledge (NEK) settings, without manual instruction engineering.
\clearpage
\begin{table*}[ht!]
\centering
\caption{\small \textbf{Case Study: Pareto Front Based Instruction Optimization.} An example optimization trajectory from \Cref{sec:mitigation-experiments}, illustrating how an initial generic instruction is progressively refined through mutation and crossover guided by gradient signals. Since the initial population contains only a single instruction ($|P_0|=1$), the crossover operator is not applicable in Round~1.}
\label{tab:case-study}
\scriptsize
\begin{tabularx}{\textwidth}{l|p{4cm}|X}
\toprule
\textbf{Round} & \textbf{Operator} & \textbf{Instruction / Gradient} \\ \midrule
\textbf{Round 0} 
& \textbf{Initial Instruction} 
& Given a web search query, retrieve relevant passages that answer the query. 
\\ 
\midrule
\textbf{Round 1} 
& \textbf{Mutation} 
& ... 
\newline 
\textbf{Instruction ($p \in P_0$):} Given a web search query, retrieve relevant passages that answer the query.
\newline
\textbf{Gradient ($g$ in \Cref{eq:estimate}):} The current prompt ``Given a web search query, retrieve relevant passages that answer the query'' may not be specific enough to guide the model to prioritize the most up-to-date and accurate information, leading to irrelevant or outdated documents being ranked higher than the positive documents. This lack of specificity can cause the model to focus on general relevance rather than the exact information being queried, resulting in incorrect rankings. 
\newline 
\textbf{Derived Instruction ($p_\text{mut}$ in \Cref{eq:mutation}):} Given a specific query, retrieve the most relevant and up-to-date passage that directly answers the question, prioritizing information that is both accurate and current as of the query timestamp.
\newline 
...
\\ 
\midrule
\multicolumn{3}{c}{\dots (Round optimization steps) \dots} \\ \midrule
\textbf{Round 4} 
& \textbf{Mutation} 
& ... 
\newline 
\textbf{Instruction ($p \in P_3$):} Given a specific query, retrieve the most relevant and up-to-date passage that directly answers the question, prioritizing information that is both accurate and current as of the query timestamp.
\newline 
\textbf{Gradient ($g$ in \Cref{eq:estimate}):} The current prompt lacks specificity regarding the handling of out-of-date or less relevant information, which might lead the model to prioritize documents based on their general relevance to the query rather than their accuracy or up-to-dateness, as indicated by the provided timestamp or other temporal cues. This could cause the model to assign higher scores to negative documents that, although less current or less accurate, contain closely related keywords or topics, thus overriding the importance of recency and factual precision in the scoring process.
\newline 
\textbf{Derived Instruction ($p_\text{mut}$ in \Cref{eq:mutation}):} Given a specific query, retrieve the most relevant and up-to-date passage that directly answers the question, prioritizing information that is both accurate and current as of the query timestamp, while considering the temporal relevance of the documents and favoring those with more recent timestamps when the query implies a need for the latest information. 
\newline 
...
\\ \cmidrule{2-3}
& \textbf{Crossover}
& ... 
\newline 
\textbf{Instruction A ($p_A$ in \Cref{eq:crossover}):} Given a specific query, retrieve the most relevant and up-to-date passage that directly answers the question, prioritizing information that is both accurate and current as of the query timestamp. 
\newline 
\textbf{Instruction B ($p_B$ in \Cref{eq:crossover}):} Given a web search query that may require the most current, specific, or nuanced information, retrieve the most relevant and up-to-date passage that directly answers the query. Ensure the information is not only accurate and reflects the latest developments or facts related to the query but also consider the context and intent behind the query to provide the most appropriate response, prioritizing understanding over mere keyword matching. 
\newline 
\textbf{Derived Instruction ($p_\text{cross}$ in \Cref{eq:crossover}):} Given a web search query that may require the most current, specific, or nuanced information, retrieve the most relevant and up-to-date passage that directly answers the question. Prioritize information that is both accurate and current as of the query timestamp. Additionally, consider the context and intent behind the query to provide the most appropriate response, ensuring the answer is not just a keyword match but a thoughtful and informed reply that reflects the latest developments or facts related to the query. 
\newline 
...
\\ 
\midrule
\multicolumn{3}{c}{\dots (Round optimization steps) \dots} \\
\bottomrule
\end{tabularx}
\end{table*}


\clearpage
\begin{table*}[h]
\centering
\small
\caption{\textbf{Performance comparison on Evolving Knowledge ($\mathcal{D}_{\text{EK}}$) and Non-Evolving Knowledge ($\mathcal{D}_{\text{NEK}}$) task.} We use RankGPT \citep{RankGPT} based on LLaMA 3.2-3B-Instruct \citep{grattafiori2024llama} as our base model. Our instruction optimization method (Pareto Solution 1-4) discovers a Pareto front offering superior trade-offs compared to baselines. The trade-off plot is provided in \Cref{appendix:pareto_plot}.}
\resizebox{1.0\textwidth}{!}{
\begin{tabular}{l|ccc|ccc}
\toprule
\multirow{2}{*}{\textbf{Method}} & \multicolumn{3}{c|}{$\mathcal{D}_{\text{EK}} \uparrow$} & \multicolumn{3}{c}{$\mathcal{D}_{\text{NEK}} \uparrow$} \\
\cmidrule(lr){2-4} \cmidrule(lr){5-7}
& MAP & MRR@10 & nDCG@10 & MAP & MRR@10 & nDCG@10 \\
\midrule
\textbf{Base Model} & 15.51 & 13.25 & 19.26 & 28.92 & 38.46 & 48.78 \\
\midrule
\textbf{Temporal-Aware Models} & & & & & & \\
\hspace{1em} TempRALM \citep{tempralm} & 19.97 & 17.78 & 23.48 & 26.08 & 31.19 & 43.31 \\
\hspace{1em} FreshPrompt \citep{freshllms} & 21.42 & 19.05 & 24.50 & 28.59 & \textbf{40.05} & 48.31 \\
\hspace{1em} MRAG \citep{mrag} & 19.40 & 17.31 & 25.16 & 28.60 & 38.12 & 48.11 \\
\midrule
\textbf{Fine Tuning} & & & & & & \\
\hspace{1em}Point-wise Finetuning & 18.34 & 16.18 & 23.73 & 28.97 & 38.99 & 48.91 \\
\hspace{1em}List-wise Finetuning & N/A & N/A & N/A & N/A & N/A & N/A \\
\midrule
\textbf{Ours} & & & & & & \\
\hspace{1em}Pareto Solution 1 & \textbf{22.28} & \textbf{20.52} & \textbf{29.67} & 28.82 & 37.91 & 48.23 \\
\hspace{1em}Pareto Solution 2 & \underline{21.87} & \underline{20.05} & \underline{29.05} & 29.31 & 38.59 & 49.19 \\
\hspace{1em}Pareto Solution 3 & 21.80 & 19.93 & 28.65 & \underline{29.68} & 39.05 & \underline{49.65} \\
\hspace{1em}Pareto Solution 4 & 21.20 & 19.25 & 27.83 & \textbf{29.71} & \underline{39.37} & \textbf{50.14} \\
\bottomrule
\end{tabular}
}
\label{tab:mitigation_2}
\end{table*}

\subsection{Pareto-Based Instruction Optimization for RankGPT (LLaMA 3.2-3B-Instruct)}
\label{appendix:mitigation_rankllama}

Beyond Qwen3-Reranker-8B, we evaluate our Pareto-based instruction optimization framework on RankGPT \citep{RankGPT}, built on LLaMA 3.2-3B-Instruct \citep{grattafiori2024llama}. 
As shown in \Cref{tab:mitigation_2}, the results exhibit a pattern consistent with \Cref{tab:mitigation}: our framework identifies a Pareto front of non-dominated solutions that improves the trade-off between $\mathcal{D}_{\text{EK}}$ and $\mathcal{D}_{\text{NEK}}$ relative to the baselines.

At one extreme, Pareto Solution 1 prioritizes $\mathcal{D}_{\text{EK}}$, achieving the highest MAP of 22.28, corresponding to a 44\% relative improvement over the base model. It outperforms strong baselines such as TempRALM and FreshPrompt on $\mathcal{D}_{\text{EK}}$, while avoiding the substantial degradation on $\mathcal{D}_{\text{NEK}}$ observed with TempRALM. At the other extreme, Pareto Solution 4 achieves the best MAP on $\mathcal{D}_{\text{NEK}}$ at 29.71. Notably, it still attains a MAP of 21.20 on $\mathcal{D}_{\text{EK}}$, remaining above the base model despite prioritizing $\mathcal{D}_{\text{NEK}}$. The intermediate solutions further exhibit a smooth trade-off between the two objectives.

\begin{wrapfigure}{r}{0.42\columnwidth}
\centering

\begin{subfigure}{0.48\linewidth}
\includegraphics[width=\linewidth]{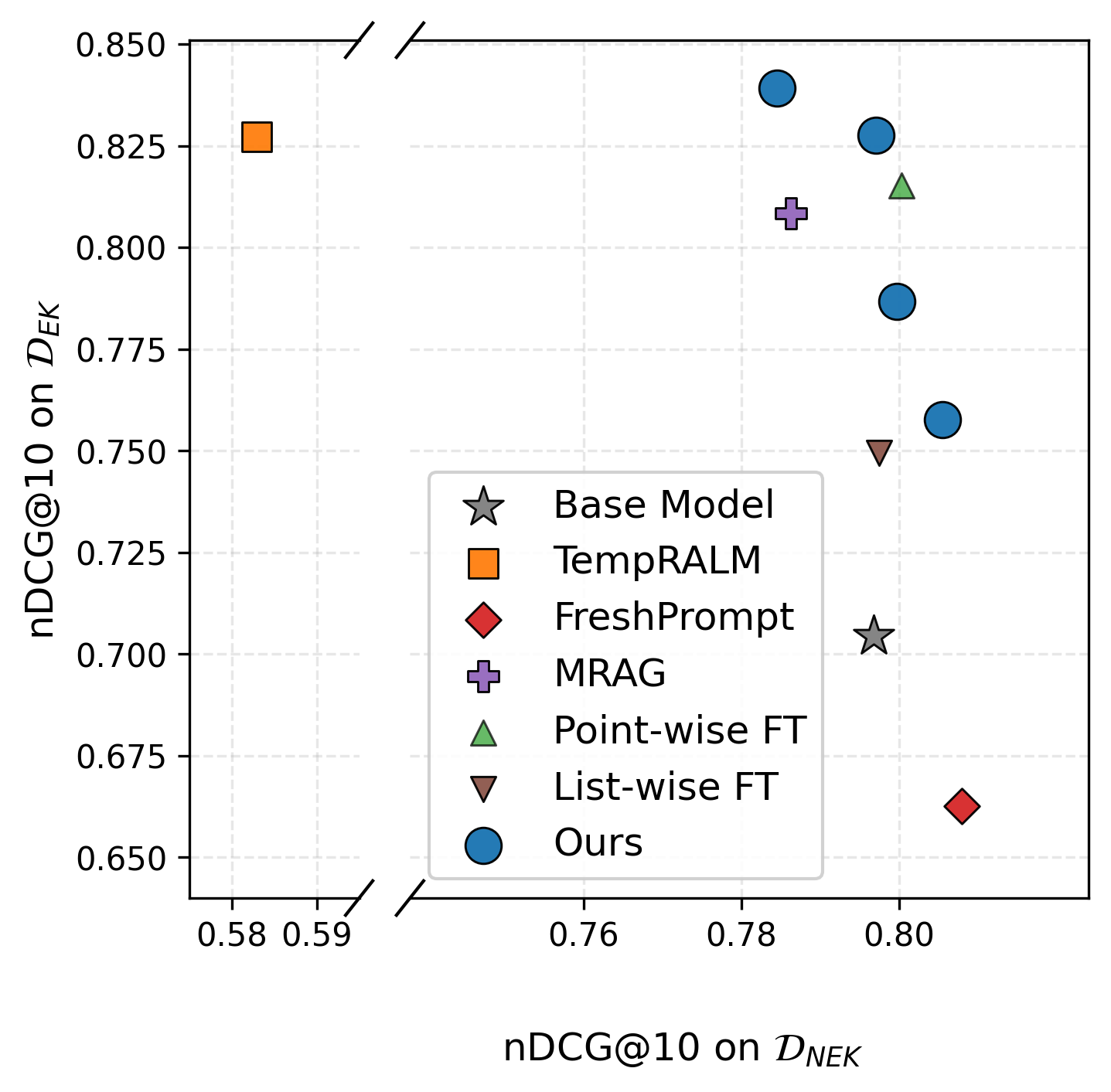}
\caption{\scriptsize Qwen3-Reranker-8B}
\label{fig:pareto_qwen}
\end{subfigure}
\hfill
\begin{subfigure}{0.46\linewidth}
\includegraphics[width=\linewidth]{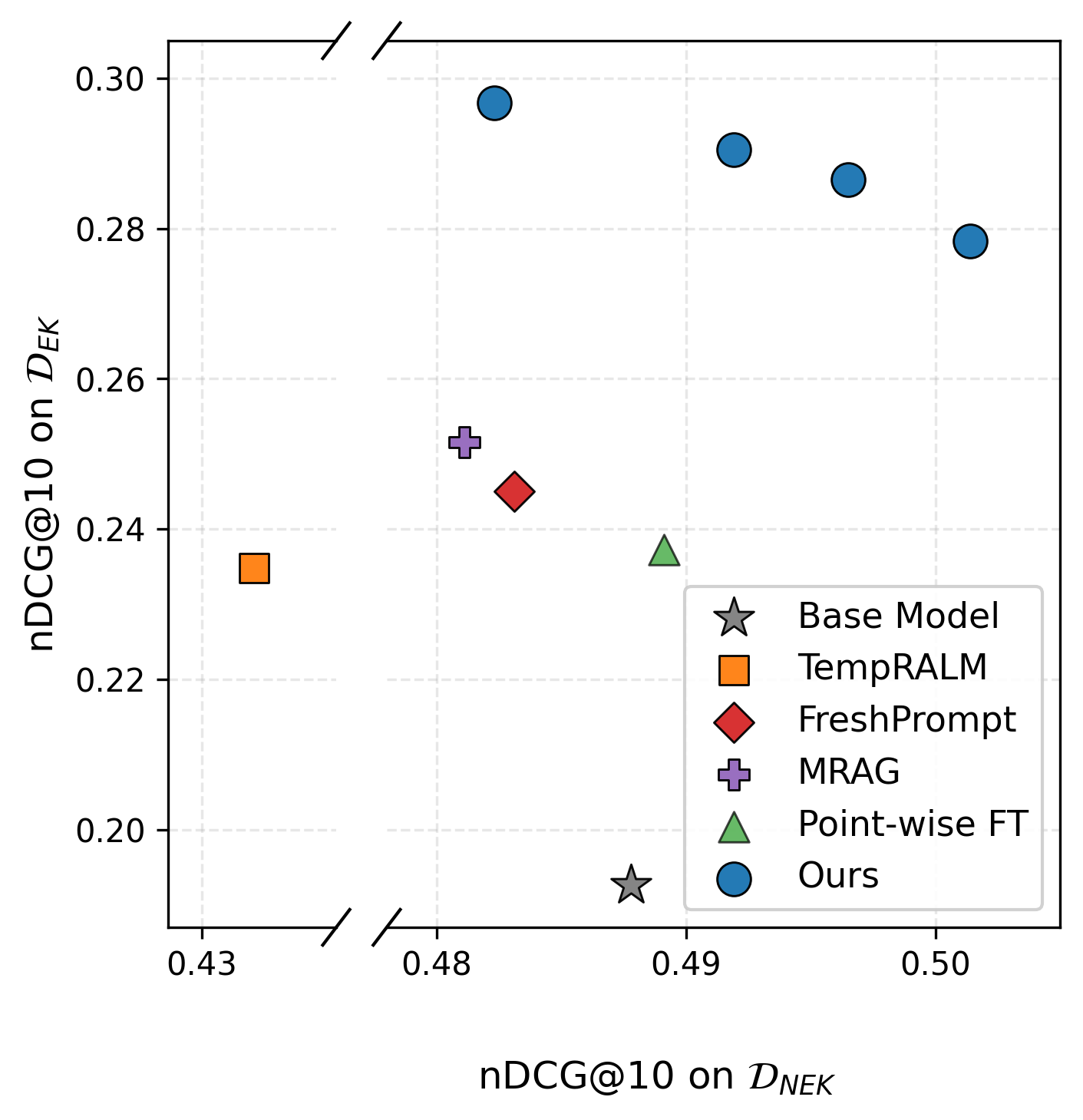}
\caption{\scriptsize RankGPT (LLaMA)}
\label{fig:pareto_rankllama}
\end{subfigure}

\caption{\textbf{Comparison of nDCG@10 scores for Qwen3-Reranker-8B and RankGPT (LLaMA 3.2-3B-Instruct).}}
\label{fig:pareto_plot}
\vspace{-10pt}
\end{wrapfigure}




For the fine-tuning baseline, we note that RankGPT is a listwise re-ranker, whereas our setting does not provide full permutation supervision over the candidate passages. We therefore adopt a point-wise fine-tuning setup, training the model to identify the most relevant passage among the candidates and thereby predict the positive passage.

\subsection{Trade-off Plots}
\label{appendix:pareto_plot}
We plotted the nDCG@10 scores of our method and the baseline methods on $\mathcal{D}_{\text{EK}}$ and $\mathcal{D}_{\text{NEK}}$ in \Cref{fig:pareto_plot}. The results show that, for both Qwen3-Reranker-8B (\Cref{tab:mitigation}) and RankGPT (LLaMA 3.2-3B-Instruct) (\Cref{tab:mitigation_2}), our method identifies a Pareto front of non-dominated solutions that achieves a superior trade-off between $\mathcal{D}_{\text{EK}}$ and $\mathcal{D}_{\text{NEK}}$ compared to the baselines.

\section{Model Details}
\label{sec:appendix_model_details}
\subsection{Re-ranker Details}
\begin{table*}[htbp]
\centering
\caption{\textbf{Detailed Model Specifications} for all re-rankers benchmarked in this work. For reference and reproducibility, we include model name, creator organization, initial release date, hosting platform, availability type, and model size.}
\label{tab:model_spec}
\small

\begin{threeparttable}
\resizebox{\textwidth}{!}{%
\begin{tabular}{@{}llllll@{}}
\toprule
\textbf{Model} & \textbf{Created by} & \textbf{Release Date} & \textbf{Hosted by} & \textbf{Availability Type} & \textbf{Model Size} \\ 
\midrule
MonoT5 \citep{monot5} & \citet{monot5} & 03/14/2020 & Huggingface & Open Weight & 220M (T5-base) \\
RankT5 \citep{rankt5} & \citet{rankt5} & 10/12/2022 & Huggingface & Open Weight & 220M (T5-base) \\
UPR \citep{upr} & \citet{upr} & 04/03/2023 & Huggingface & Open Weight & 220M (T5-base) \\
RankGPT \citep{RankGPT} & \citet{RankGPT} & 04/19/2023 & OpenAI & Proprietary & N/A \\
RankVicuna \citep{Rankvicuna} & \citet{Rankvicuna} & 09/26/2023 & Huggingface & Open Weight & 7B \\
RankZephyr \citep{Rankzephyr} & \citet{Rankzephyr} & 12/05/2023 & Huggingface & Open Weight & 7B \\
bce-reranker-base-v1 \citep{youdao_bcembedding_2023} & NetEase Youdao & 01/03/2024 & Huggingface & Open Weight & 279M \\
InRanker \citep{inranker} & \citet{inranker} & 01/12/2024 & Huggingface & Open Weight & 220M \\
mxbai-rerank-base-v1 \citep{rerank2024mxbai} & Mixedbread & 02/29/2024 & Huggingface & Open Weight & 184M \\
Twolar \citep{twolar} & \citet{twolar} & 03/26/2024 & Huggingface & Open Weight & 0.7B (twolar-large) \\
jina-reranker-v1-tiny-en \tnote{a} & Jina AI & 04/15/2024 & Huggingface & Open Weight & 33M \\
jina-reranker-v1-turbo-en \tnote{b} & Jina AI & 04/15/2024 & Huggingface & Open Weight & 37.8M \\
jina-reranker-v2-base-multilingual \tnote{c} & Jina AI & 06/19/2024 & Huggingface & Open Weight & 0.3B \\
gte-multilingual-reranker-base \citep{zhang2024mgte} & Alibaba & 07/20/2024 & Huggingface & Open Weight & 306M \\
LdlR-Qwen2-reranker-1.5B & neofung \tnote{d} & 08/12/2024 & Huggingface & Open Weight & 1.5B \\ 
IncontextReranker \citep{incontext_reranker} & \citet{incontext_reranker} & 10/03/2024 & Huggingface & Open Weight & 8B (Llama-3.1-8B) \\ 
Qwen3-Reranker-0.6B \citep{Qwen3} & Alibaba & 05/29/2025 & Huggingface & Open Weight & 0.6B \\
Qwen3-Reranker-8B \citep{Qwen3} & Alibaba & 05/29/2025 & Huggingface & Open Weight & 8B \\ 
\bottomrule
\end{tabular}
}
\begin{tablenotes}\footnotesize
\item[a] \url{https://jina.ai/models/jina-reranker-v1-tiny-en/}
\item[b] \url{https://jina.ai/models/jina-reranker-v1-turbo-en/}
\item[c] \url{https://jina.ai/models/jina-reranker-v2-base-multilingual/}
\item[d] the user account on Hugging Face
\end{tablenotes}
\end{threeparttable}

\end{table*}
\Cref{tab:model_spec} summarizes the detailed specifications of all re-rankers
evaluated in our benchmark.
For each model, we report the creator organization, initial release date,
hosting platform, availability type (\textit{e.g.}, open-weight or proprietary),
and model size.

\subsection{Listwise Re-ranking Prompts}

To ensure reproducibility, we provide the exact prompt templates used for listwise re-rankers.

\paragraph{Prompt Template for RankGPT}
\begin{promptbox}[]
system: You are RankGPT, an intelligent assistant that can rank passages based on their relevancy to the query.
user: I will provide you with 20 passages, each indicated by number identifier []. 
Rank the passages based on their relevance to query: {query}.
assistant: Okay, please provide the passages.
user: [1] {candidate document 1}
assistant: f"Received passage [1].
user: [2] {candidate document 2}
assistant: f"Received passage [2].
...
assistant: f"Received passage [20].
user: Search Query: {query}.
Rank the 20 passages above based on their relevance to search query. The passages should be listed in descending order using identifiers. The most relevant passages should be listed first. The output format should be [] > [], e.g., [1] > [2]. Only response the ranking results, do not say any word or explain.
\end{promptbox}

\paragraph{Prompt Template for RankVicuna and RankZephyr}
\begin{promptbox}[]
A chat between a curious user and an artificial intelligence assistant.
The assistant gives helpful, detailed, and polite answers to the user's questions.

USER:
I will provide you with 20 passages, each indicated by a numerical identifier [].
Rank the passages based on their relevance to the search query: {query}

[1] {candidate document 1}
[2] {candidate document 2}
...
[20] {candidate document 20}

Search Query: {query}

Rank the 20 passages above based on their relevance to the search query.
All the passages should be included and listed using identifiers,
in descending order of relevance.

The output format should be [] > [], e.g., [4] > [2].
Only respond with the ranking results, do not say any word or explain.

ASSISTANT:
\end{promptbox}

\section{Dataset Details}
The final \Dataset{} benchmark comprises 3,658 queries for each query type (explicitly and implicitly recency-seeking), totaling 7,316 unique queries. Each query is associated with a candidate set containing 51 passages.
For our benchmarking experiments evaluating existing re-rankers, we utilize the entire dataset.
For the experiments detailed in \Cref{sec:mitigation}, this dataset was partitioned into three subsets: training, validation, and testing. Specifically, we employed 200 instances for training, 800 instances for validation, and reserved the remaining portion for the test set.
All benchmarking experiments were run on four NVIDIA A100 80GB GPUs, with each re-ranker evaluated in less than two hours.

\section{Human Annotation Details}
\label{sec:appendix_human_eval}
\begin{figure*}[t!]
\centering
    \includegraphics[width=1.0\linewidth]{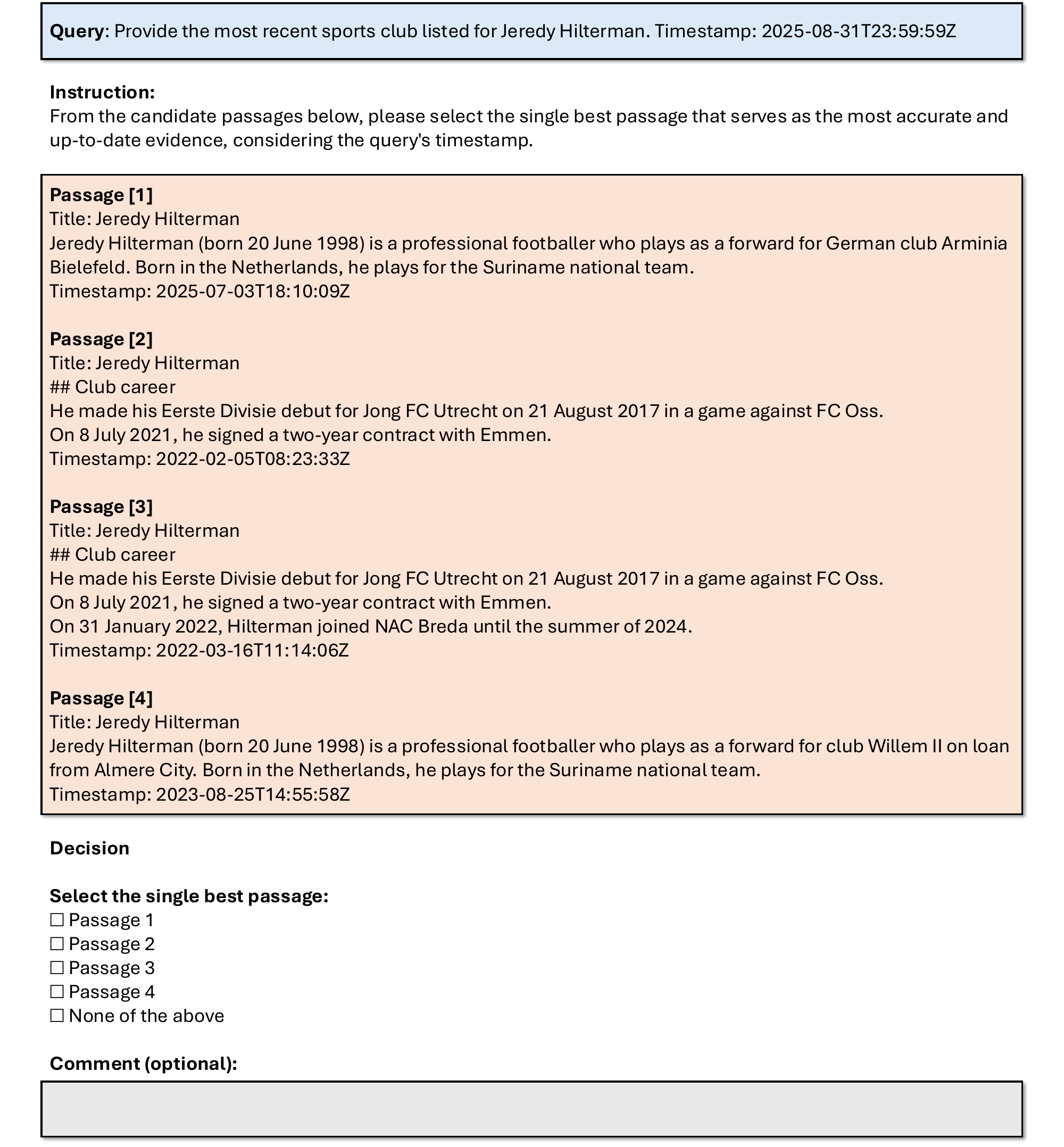}
    \caption{\small \textbf{Example of the human annotation interface.} Annotators were presented with a query (including timestamp) and four passages (one positive, three hard negatives) labeled by our pipeline. They were asked to select the single passage that provided sufficient evidence for the query and was temporally valid given the query timestamp, or select \texttt{None of the above} if no passage met both criteria.}
    \label{fig:annotation_example}
\end{figure*}

To rigorously assess the quality of the automatically constructed \Dataset{} benchmark, we performed a human evaluation study.

\paragraph{Setup}
We randomly sampled 200 queries from the benchmark. For each query, we presented three trained human annotators with a curated set of four passages. This set consisted of:
\begin{itemize}
    \item The single passage designated as \textbf{positive} by our dataset construction pipeline.
    \item The three \textbf{highest-scoring hard negative passages} as ranked by the Qwen3-Embedding-0.6B model during the final assembly phase (see \Cref{sec:dataset_construction}). 
\end{itemize}
Presenting only the most challenging negatives focused the annotators' efforts on the most ambiguous cases, ensuring an efficient and targeted validation process.

\paragraph{Annotation Task}
The annotators were instructed to carefully read the query (including its associated timestamp) and all four candidate passages. Their task was to select the \emph{single passage} that was \emph{most temporally aligned} with the query timestamp and provided \emph{sufficient and correct evidence} to answer the query. An example of the annotation interface and task is shown in \Cref{fig:annotation_example}. To handle potential imperfections in our automated pipeline or inherent ambiguities in the source material, annotators were given the option to select \texttt{None of the above} if they determined that none of the provided passages met both criteria. The annotators were informed that their judgments would be used solely for research and evaluation purposes, and they participated on a voluntary basis.

\paragraph{Evaluation Metrics}
We employed two standard metrics to quantify the quality of our dataset based on the human annotations:

\begin{enumerate}
    \item \textbf{Inter-Annotator Agreement (IAA):} We measured the consistency among the three annotators using Fleiss' Kappa ($\kappa$). This metric assesses the reliability of agreement beyond chance. A high $\kappa$ value indicates that the annotation task was clear and the judgments were consistent.
    \item \textbf{Agreement with Pipeline Labels:} This metric directly validates our automatic labeling process. We compared the majority vote label from the three annotators (\textit{i.e.}, the label agreed upon by at least two annotators) against the original positive label assigned by our pipeline. A high agreement rate signifies that our pipeline accurately identifies the correct passage.
\end{enumerate}

\paragraph{Results and Disagreement Analysis}
Our analysis yielded a \emph{Fleiss' Kappa score of $\kappa = 0.9689$}. According to standard interpretations, this value represents \emph{almost perfect agreement}, confirming the clarity of the task and the reliability of our annotators' judgments.
The human majority vote aligned with our pipeline's designated positive label on \emph{197 out of the 200} sampled queries. This corresponds to a high \emph{agreement rate of 98.5\%}.
We manually reviewed the 3 instances where the human majority vote disagreed with the pipeline label. In all such cases, the discrepancy arose because none of the presented passages, including the one designated as positive by our pipeline, offered sufficiently clear or direct evidence to definitively answer the query. For instance, a query asking for the most recent team an entity was \emph{coaching} might be paired with passages only mentioning the team the entity last \emph{played} for. Likewise, a query about a team's latest head coach might yield passages discussing the team's roster or recent performance without explicitly naming the coach. These few disagreements primarily highlight the inherent challenge of finding perfectly explicit textual evidence for every fact.

\paragraph{Conclusion}
The high inter-annotator agreement and the strong agreement rate between human judgments and our pipeline labels provide robust validation for the quality of \Dataset{}. The analysis confirms its suitability as a benchmark for evaluating re-rankers under temporally evolving information.

\section{Implementation Details}
\label{sec:appendix_implementation_details}

\subsection{Evaluation Metrics}
\label{sec:appendix_eval_metrics}

We evaluate the performance of re-ranking models using standard Information Retrieval (IR) metrics to assess the quality of the generated ranking $\pi = f_{\theta}(q, \mathcal{C}_q; p)$ relative to ground-truth relevance labels $\mathcal{R}_q$. Here, $p$ denotes an optional instruction that can be provided to the model. Let $\pi(i)$ denote the passage ranked at position $i$ ($i=1$ being the top rank) within the set of $m = |\mathcal{C}_q|$ candidates. We define $\mathcal{C}_q^+ \subseteq \mathcal{C}_q$ as the subset of positive passages for query $q$ based on $\mathcal{R}_q$.

\paragraph{Mean Average Precision (MAP).}
Average Precision (AP) measures the precision of a model integrated over the recall curve, effectively rewarding models that rank positive documents at higher positions. Precision at rank $k$ ($P@k$) is defined as:
$$
P@k = \frac{|\{ i \mid 1 \le i \le k, \pi(i) \in \mathcal{C}_q^+ \}|}{k}.
$$
The AP for a single query $q$ is calculated as:
$$
AP(q) = \frac{1}{|\mathcal{C}_q^+|} \sum_{k=1}^{m} \left( P@k \cdot \mathbbm{1}(\pi(k) \in \mathcal{C}_q^+) \right),
$$
where $\mathbbm{1}(\cdot)$ is the indicator function. If $|\mathcal{C}_q^+| = 0$, we set $AP(q) = 0$. MAP is the arithmetic mean of AP scores across the set of queries $Q$:
$$
MAP = \frac{1}{|Q|} \sum_{q \in Q} AP(q).
$$

\paragraph{Mean Reciprocal Rank (MRR).}
Reciprocal Rank (RR) focuses on the rank of the first positive document. Let $r_1$ be the rank of the first positive passage in $\pi$:
$$
r_1 = \min \{ i \mid 1 \le i \le m, \pi(i) \in \mathcal{C}_q^+ \}.
$$
If no positive passage exists in $\pi$, we define $RR(q) = 0$; otherwise, $RR(q) = 1/r_1$. MRR is the mean of RR scores across all queries $Q$. When reported at a specific cutoff $k$ (MRR@$k$), $RR(q)$ is set to 0 if $r_1 > k$.

\paragraph{Normalized Discounted Cumulative Gain (nDCG).}
nDCG evaluates the gain of a document based on its position, applying a logarithmic discount at lower ranks. Under our binary relevance setting (gain $G(i) = 1$ if $\pi(i) \in \mathcal{C}_q^+$ and $0$ otherwise), the DCG at rank $k$ is:
$$
DCG@k = \sum_{i=1}^{k} \frac{G(i)}{\log_2(i+1)}.
$$
To allow for cross-query comparison, DCG is normalized by the Ideal DCG (IDCG@$k$), which represents the maximum possible DCG achievable by ranking all positive documents at the top:
$$
nDCG@k = \frac{DCG@k}{IDCG@k}.
$$
If $IDCG@k = 0$, we define $nDCG@k = 0$.

\paragraph{Hit Rate.}
Hit Rate at rank $k$ is a binary metric that indicates whether at least one positive passage is present within the top $k$ positions of the ranked list. It is particularly relevant in scenarios where the user's information need is satisfied by finding any single positive item. Using the previously defined rank of the first relevant passage $r_1$, the Hit Rate for a query $q$ is defined as:
$$
Hit Rate@k(q) = \mathbbm{1}(r_1 \le k),
$$
where $\mathbbm{1}(\cdot)$ is the indicator function. If no positive passage exists in $\pi$, $Hit Rate@k(q) = 0$. The Mean Hit Rate@$k$ is the average of these scores across all queries in $Q$.


\subsection{Baselines}
\label{sec:appendix_baselines}

We provide further details on the baseline methods compared in our experiments (\Cref{sec:mitigation}).

\subsubsection{Base Model}
We use the re-ranker's default instruction as provided in the original work \citep{Qwen3}, which is typically a general instruction like ``Given a web search query, retrieve relevant passages that answer the query''. This baseline represents the standard, non-adapted performance of a powerful contemporary re-ranker on our tasks.

\subsubsection{Temporal-Aware Models}
We compare against three recent methods specifically designed to improve the temporal awareness of information retrieval or RAG systems.

\paragraph{TempRALM}
TempRALM \citep{tempralm} augments the standard semantic relevance score $s(q, d)$ between a query $q$ and a document $d$ with a temporal relevance score $r(q_t, d_t)$. Here, $q_t$ is the query timestamp and $d_t$ is the document timestamp. The temporal score $r(q_t, d_t)$ is inversely proportional to the time difference $|q_t - d_t|$, encouraging the re-ranking of documents temporally closer to the query timestamp. The final retrieval score is calculated as: $s(q, d) + r(q_t, d_t)$.
We adapt this concept by applying the scoring mechanism to the re-ranking candidates provided to the base re-ranker model to assess its impact in a re-ranking context.

\paragraph{FreshPrompt}
FreshPrompt \citep{freshllms} prepends the main prompt with few-shot demonstrations. Each demonstration typically includes an example query, a set of evidence snippets. Although the original format focuses on generation, we adapt the principle for re-ranking: we provide the re-ranker with demonstrations showcasing how to prioritize temporally relevant passages from a candidate list containing both positive and negative examples relative to a query timestamp.

\paragraph{MRAG}
MRAG \citep{mrag} employs a modular framework involving three key stages:
\begin{enumerate}
    \item \textbf{Question and Passage Processing:} The input query is segmented into its main content (MC) and temporal constraints (TC). In addition, each passage is summarized with a LLM (LLaMA-3.3-70B-Inst here).
    \item \textbf{Semantic-Temporal Hybrid Ranking:} A final ranking module multiplicatively combines semantic scores with symbolic temporal scores derived using temporal score functions similar to the temporal activation functions in \citet{CHEN2022109134}.
\end{enumerate}
We apply the principles of MRAG's hybrid ranking logic within our re-ranking evaluation framework.

\subsubsection{Fine-Tuning}
To establish strong supervised baselines, we fine-tune the Qwen3-Reranker-8B base model on our task data. The fine-tuning process utilizes a combined training set containing examples from both $\mathcal{D}_{\text{EK}}$ and $\mathcal{D}_{\text{NEK}}$. We employ two standard learning-to-rank approaches:
\begin{itemize}
    \item \textbf{Point-wise Fine-tuning:} Trains the model to predict the binary relevance label (relevant/irrelevant) for each query-passage pair independently, typically using a cross-entropy loss.
    \item \textbf{List-wise Fine-tuning:} Trains the model to optimize the order of the entire list of passages for a given query, often using losses that directly approximate ranking metrics like ListNet or LambdaRank.
\end{itemize}
We follow standard procedures outlined in the official repository\footnote{\url{https://github.com/QwenLM/Qwen3-Embedding/blob/main/docs/training/SWIFT.md}} for fine-tuning the re-ranker. These baselines represent a resource-intensive but potentially powerful approach for adapting the model to our scenario, serving as a contrast to our parameter-free instruction optimization technique.

\subsection{Algorithm}
\label{sec:appendix_algo}
\begin{algorithm*}[t]
\caption{Pareto-Based Instruction Optimization via Evolutionary Search}
\label{alg:moi_optimization}
\KwIn{Initial population $P^{(0)}$, Datasets $\mathcal{D}_{\text{EK}}, \mathcal{D}_{\text{NEK}}$, Max Pareto size $B$, Rounds $T$, Expansion factor $E$, Mutation $(\mathcal{G}_{\text{est}}, \mathcal{G}_{\text{apply}})$, Crossover $\mathcal{X}$, Utility $\mathcal{U}$}
\KwOut{Optimized Pareto-front set $P^{(T)}$}

\For{$t = 0$ \KwTo $T-1$}{
    $P_{\text{mut}} \gets \emptyset$; $P_{\text{cross}} \gets \emptyset$\;
    
    \tcp{Expansion Phase: Generate candidates via Mutation and Crossover}
    Sample training minibatch $\mathcal{B}_{\text{tr}} = \mathcal{B}^{\text{tr}}_{\text{EK}} \cup \mathcal{B}^{\text{tr}}_{\text{NEK}}$\;
    
    \tcp{1. Mutation}
    \ForEach{$p \in P^{(t)}$}{
        $\mathcal{E}(p) \gets \{ (q, c^*, \mathcal{E}_q(p)) \mid (q, \mathcal{C}_q, \mathcal{R}_q) \in \mathcal{B}_{\text{tr}}, \mathcal{E}_q(p) \neq \emptyset \}$\;
        \hfill where $\mathcal{E}_q(p) = \{ c' \in \mathcal{C}_q \setminus \{c^*\} : \pi_{p,q}(c') < \pi_{p,q}(c^*) \}$\;
            
        $\mathbf{g} \gets \mathcal{G}_{\text{estimate}}(p, \mathcal{E}(p))$ \tcp*[r]{Estimate textual gradients}
        
        \ForEach{$g \in \mathbf{g}$}{
             $p_{\text{mut}} \gets \mathcal{G}_{\text{apply}}(p, \mathcal{E}(p), g)$\ \tcp*[r]{Apply textual gradients}
             $P_{\text{mut}} \gets P_{\text{mut}} \cup \{p_{\text{mut}}\}$\;
        }
    }

    \tcp{2. Crossover}
    Select parent pairs $\{(p_A, p_B)\}$ from $P^{(t)}$ to satisfy expansion factor $E$\;
    \ForEach{pair $(p_A, p_B)$}{
        $\mathcal{E}_{A \succ B} \gets \{ \text{examples in } \mathcal{B}_{\text{tr}} \text{ s.t. } \pi_{p_A, q}(c^*) < \pi_{p_B, q}(c^*) \}$\;
        $\mathcal{E}_{B \succ A} \gets \{ \text{examples in } \mathcal{B}_{\text{tr}} \text{ s.t. } \pi_{p_B, q}(c^*) < \pi_{p_A, q}(c^*) \}$\;
        
        $P_{\text{new}} \gets \mathcal{X}(p_A, p_B, \mathcal{E}_{A \succ B}, \mathcal{E}_{B \succ A})$ \tcp*[r]{Synthesize hybrid prompts}
        $P_{\text{cross}} \gets P_{\text{cross}} \cup P_{\text{new}}$\;
    }
    
    $P_{\text{cand}} \gets P^{(t)} \cup P_{\text{mut}} \cup P_{\text{cross}}$\;

    \tcp{Evaluation Phase: Multi-objective Scoring}
    Sample validation minibatch $\mathcal{B}_{\text{val}} = \mathcal{B}^{\text{val}}_{\text{EK}} \cup \mathcal{B}^{\text{val}}_{\text{NEK}}$\;
    \ForEach{$p \in P_{\text{cand}}$}{
        $\hat{\mathcal{J}}_{\text{EK}}(p) \gets \frac{1}{|\mathcal{B}^{\text{val}}_{\text{EK}}|} \sum \mathcal{U}(f_{\theta}(q, \mathcal{C}_q; p), \mathcal{R}_q)$\;
        $\hat{\mathcal{J}}_{\text{NEK}}(p) \gets \frac{1}{|\mathcal{B}^{\text{val}}_{\text{NEK}}|} \sum \mathcal{U}(f_{\theta}(q, \mathcal{C}_q; p), \mathcal{R}_q)$\;
        $\hat{\mathbf{F}}(p) \gets (\hat{\mathcal{J}}_{\text{EK}}(p), \hat{\mathcal{J}}_{\text{NEK}}(p))$\;
    }

    \tcp{Selection Phase: Pareto Pruning}
    Define $p \prec p' \iff (\hat{\mathbf{F}}(p') \ge \hat{\mathbf{F}}(p)) \land (\hat{\mathbf{F}}(p') \neq \hat{\mathbf{F}}(p))$\;
    $P_{\text{front}} \gets \{ p \in P_{\text{cand}} \mid \nexists\, p' \in P_{\text{cand}} \text{ s.t. } p \prec p' \}$\;

    \lIf{$|P_{\text{front}}| > B$}{
        $P^{(t+1)} \gets \text{SelectTopByCrowding}(P_{\text{front}}, B)$
    }
    \lElse{
        $P^{(t+1)} \gets P_{\text{front}}$
    }
}
\Return $P^{(T)}$
\end{algorithm*}
We provide a detailed description of our instruction optimization procedure in \Cref{alg:moi_optimization}. The algorithm outlines the iterative process of expanding the instruction population through mutation and crossover, evaluating candidate instructions against both EK and NEK objectives, and selecting the non-dominated solutions to form the Pareto front for the subsequent round. This evolutionary search effectively navigates the discrete instruction space to discover instructions that offer superior trade-offs.

In \Cref{sec:mitigation-experiments}, our evolutionary search algorithm is configured to run for 10 rounds. In each round, the total number of expanded instructions is set to $8 \cdot |P_t|$, where $|P_t|$ is the size of the current instruction population. This expansion budget is allocated equally between the Mutation and Crossover operators. Following the evaluation and selection phase, the Pareto front is pruned to a maximum size of 4 to form the population for the next round. The model used to implement the textual gradient operators ($\mathcal{G}_{\text{estimate}}$, $\mathcal{G}_{\text{apply}}$) and the crossover operator ($\mathcal{X}$) is LLaMA-3.3-70B-Instruct \citep{grattafiori2024llama}. All results are reported as the average of three independent experimental runs.


\end{document}